\documentclass[pdflatex,sn-mathphys-num]{sn-jnl}

\usepackage{graphicx}%
\usepackage{multirow}%
\usepackage{amsmath,amssymb,amsfonts}%
\usepackage{amsthm}%
\usepackage{mathrsfs}%
\usepackage[title]{appendix}%
\usepackage{xcolor}%
\usepackage{textcomp}%
\usepackage{manyfoot}%
\usepackage{booktabs}%
\usepackage{algorithm}%
\usepackage{algorithmicx}%
\usepackage{algpseudocode}%
\usepackage{listings}%
\usepackage[caption=false]{subfig}

\theoremstyle{thmstyleone}%
\theoremstyle{thmstyletwo}%

\theoremstyle{thmstylethree}%

\begin{document}


\title[Article Title]{A Study of Anatomical Priors for Deep Learning-Based Segmentation of Pheochromocytoma in Abdominal CT}

\author[1]{\fnm{Tanjin Taher} \sur{Toma}}

\author[1]{\fnm{Tejas Sudharshan} \sur{Mathai}}

\author[6]{\fnm{Bikash} \sur{Santra}} 

\author[1]{\fnm{Pritam} \sur{Mukherjee}}

\author[1]{\fnm{Jianfei} \sur{Liu}}

\author[5]{\fnm{Wesley} \sur{Jong}}

\author[5]{\fnm{Darwish} \sur{Alabyad}}

\author[1]{\fnm{Vivek} \sur{Batheja}}

\author[2]{\fnm{Abhishek} \sur{Jha}}

\author[3]{\fnm{Mayank} \sur{Patel}}

\author[1]{\fnm{Darko} \sur{Pucar}}

\author[3]{\fnm{Jayadira del} \sur{Rivero}}

\author[2,7]{\fnm{Karel} \sur{Pacak}}

\author*[1]{\fnm{Ronald M.} \sur{Summers}}\email{rms@nih.gov}

\affil[1]{\orgdiv{Imaging Biomarkers and Computer-Aided Diagnosis Laboratory}, \orgname{Clinical Center, NIH}, \orgaddress{\city{Bethesda}, \state{MD}, \country{USA}}}

\affil[2]{\orgname{Eunice Kennedy Shriver National Institute of Child Health and Human Development, NIH}, \orgaddress{\city{Bethesda}, \state{MD}, \country{USA}}} 

\affil[3]{\orgname{National Cancer Institute, NIH}, \orgaddress{\city{Bethesda}, \state{MD}, \country{USA}}} 

\affil[4]{\orgname{National Institute of Diabetes and Digestive and Kidney Diseases, NIH}, \orgaddress{\city{Bethesda}, \state{MD}, \country{USA}}} 


\affil[5]{\orgdiv{Department of Radiology}, \orgname{The George Washington University School of Medicine and Health Sciences}, \orgaddress{\city{Washington}, \state{DC}, \country{USA}}}

\affil[6]{\orgdiv{School of AI and Data Science}, \orgname{Indian Institute of Technology Jodhpur}, \orgaddress{\state{Rajasthan}, \country{India}}}

\affil[7]{\orgdiv{Center for Adrenal Endocrine Tumors}, \orgname{AKESO}, \orgaddress{\city{Prague 5}, \country{Czech Republic}}}





\abstract{\textcolor{red}{mention author names}}

\abstract{\textbf{Purpose:} 
Accurate segmentation of pheochromocytoma (PCC) tumors from abdominal CT scans is critical for precise tumor burden measurement, prognostic assessment and treatment planning. It may also aid in early identification of tumor genetic clusters, potentially reducing reliance on costly genetic testing. This study introduces a systematic evaluation of diverse anatomical priors to identify prior configurations that most effectively enhance deep learning-based PCC segmentation accuracy.

\textbf{Methods:} 
We employed the nnU-Net framework to evaluate eleven annotation strategies for accurate 3D segmentation of pheochromocytoma, introducing a set of novel multi-class schemes based on organ-specific anatomical priors. These priors were derived from adjacent organs commonly surrounding adrenal tumors (e.g., liver, spleen, kidney, aorta, adrenal gland, and pancreas), and were compared against a broad body-region prior used in previous work. 
The framework was trained and tested on 105 portal venous phase contrast-enhanced CT scans from 91 patients in a retrospective cohort imaged at the NIH Clinical Center.
Performance was assessed using Dice Similarity Coefficient (DSC), Normalized Surface Distance (NSD), and instance-wise F1 score.

\textbf{Results:} Among all strategies, the Tumor + Kidney + Aorta (TKA) annotation yielded the highest segmentation accuracy, significantly outperforming the previously used Tumor + Body (TB) annotation across DSC (\textit{p} = 0.0097), NSD (\textit{p} = 0.0110), and F1 score (25.84\% improvement at an intersection-over-union (IoU) threshold of 0.5), measured on the test set using a 70\%-30\% train–test split. The TKA model also demonstrated superior tumor burden quantification (\textit{R}\textsuperscript{2} = 0.968), as well as strong segmentation performance across all genetic subtypes. Furthermore, when evaluated on the full dataset using five-fold cross-validation, TKA consistently outperformed TB across a range of IoU thresholds (0.1 to 0.5) in the PCC tumor detection task, reinforcing its robustness and generalizability.

%
\textbf{Conclusion:} We developed a 3D deep learning-based approach for accurate PCC tumor segmentation from contrast-enhanced CT scans, leveraging relevant spatial context from surrounding organs. The approach demonstrated high accuracy in delineating tumor boundaries, reliably detecting individual tumor instances, and quantifying tumor burden with precision, offering a valuable tool to support clinical assessment and longitudinal disease monitoring in PCC patients.
}




\maketitle

\newpage
\section{Introduction}
\label{sec1}

\begin{figure}[htbp]
  \centering
  \resizebox{0.85\textwidth}{!}{
    \begin{minipage}{\textwidth}
      \centering

      \begin{minipage}[b]{0.45\textwidth}
        \centering
        \includegraphics[width=\linewidth]{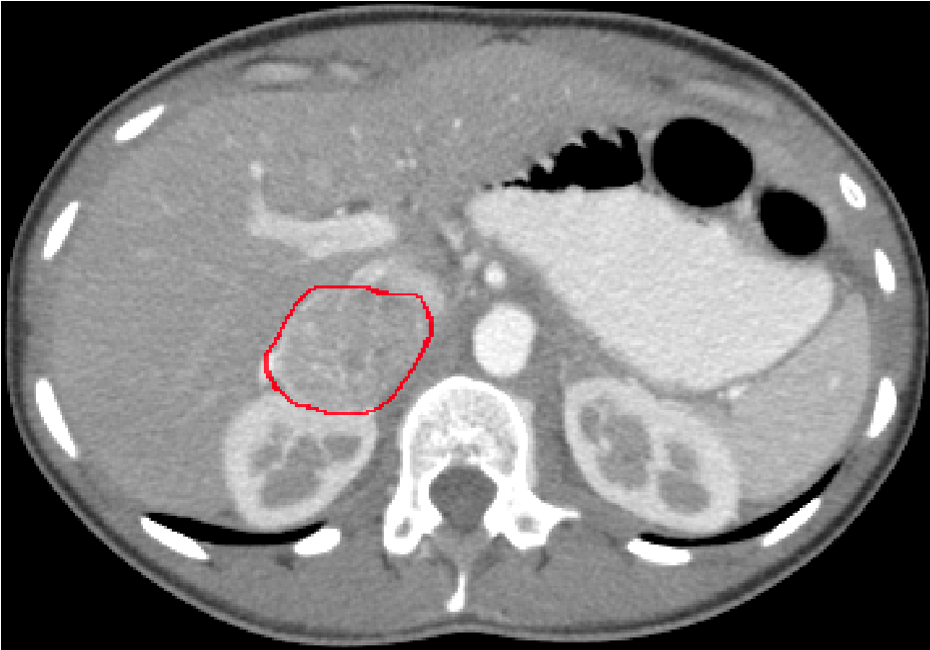}
        \par (a) \textit{SDHx}
      \end{minipage}
      \hfill
      \begin{minipage}[b]{0.45\textwidth}
        \centering
        \includegraphics[width=\linewidth]{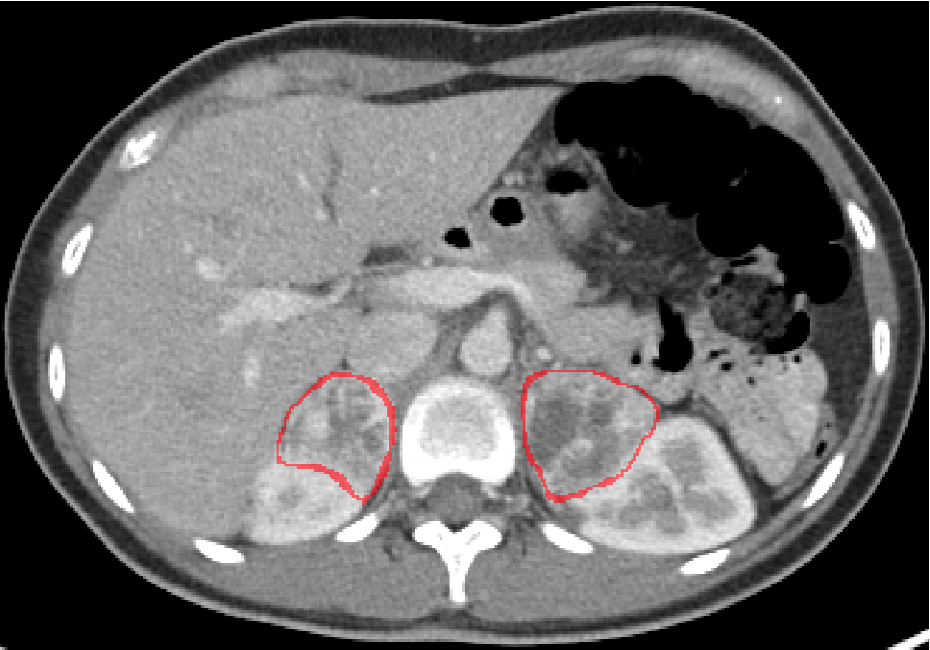}
        \par (b) \textit{VHL/EPAS1}
      \end{minipage}

      \vspace{4mm} 

      \begin{minipage}[b]{0.45\textwidth}
        \centering
        \includegraphics[width=\linewidth]{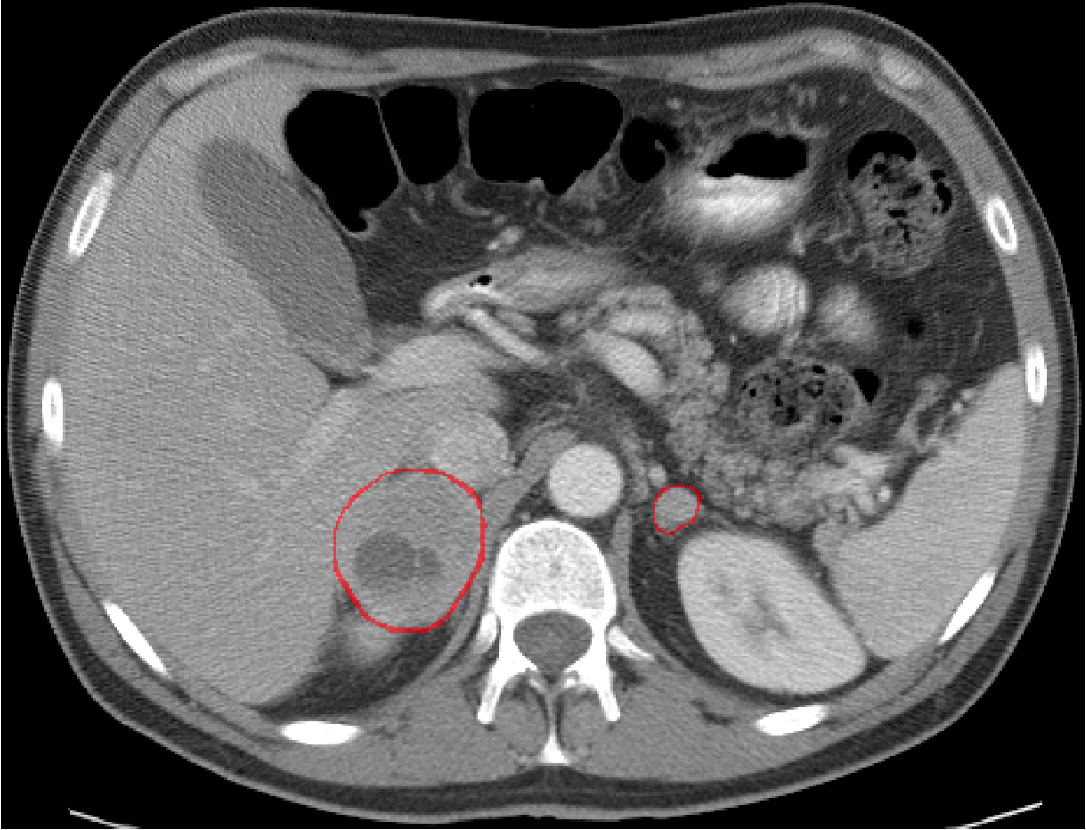}
        \par (c) Kinase
      \end{minipage}
      \hfill
      \begin{minipage}[b]{0.45\textwidth}
        \centering
        \includegraphics[width=\linewidth]{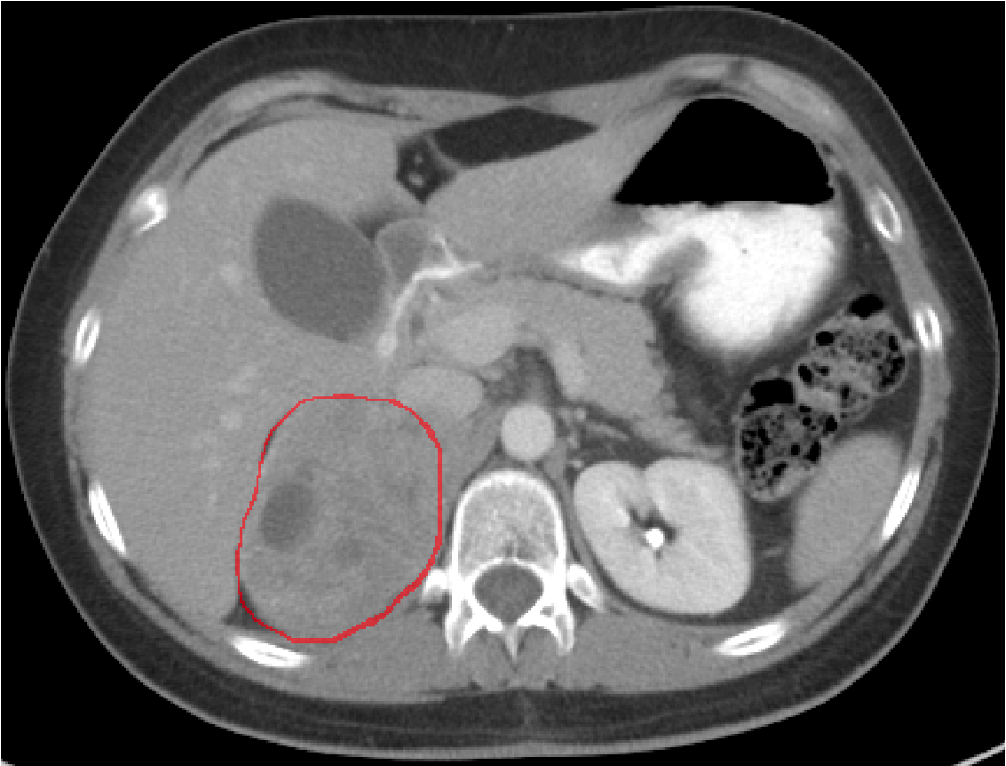}
        \par (d) Sporadic
      \end{minipage}
    \end{minipage}
  }

  \caption{Portal venous phase abdominal CT scans showing  pheochromocytoma tumors from different genetic subtypes. Red contours indicate the tumor locations in each scan. (b) and (c) show bilateral tumors.}
  \label{fig:intro_pcc_types}
\end{figure}

Pheochromocytoma (PCC) is a rare neuroendocrine tumor arising primarily from the chromaffin cells of the adrenal medulla~\cite{mercado2018pheochromocytoma}. Characterized by excessive secretion of catecholamines, pheochromocytomas can lead to significant cardiovascular morbidity, including severe hypertension, palpitations, and potentially life-threatening hypertensive crises~\cite{lenders2005phaeochromocytoma}. Besides, about 10-15\% of PCCs are metastatic~\cite{amar2005year,patel2020update}. These tumors are unique in their highly variable molecular landscape driven by genetic alterations, either germline or somatic~\cite{timmers2024imaging}. These mutations translate into three major genetic clusters: pseudohypoxia-related cluster 1A-\textit{SDHx} and cluster 1B-\textit{VHL/EPAS1} mutations, cluster 2-kinase signaling (such as \textit{RET}, \textit{NF1}, \textit{TMEM127} and \textit{MAX}), and a less defined group Wnt signaling–related cluster 3, often comprising sporadic cases~\cite{nolting2022personalized,cascon2023genetic,luca2023three}. The underlying genetic cluster can influence tumor location, biochemical profile, metastatic potential, and overall clinical management strategy~\cite{martucci2014pheochromocytoma,jain2020pheochromocytoma}. 
Timely and accurate diagnosis, localization, and characterization of these tumors are crucial for optimal management, including surgical resection, which remains the primary therapeutic intervention~\cite{strajina2017surgical,aygun2020pheochromocytoma}. Computed tomography (CT), especially contrast-enhanced abdominal imaging, is routinely employed due to its high spatial resolution, wide availability, and rapid imaging capability, making it the imaging modality of choice for initial detection and characterization of adrenal masses~\cite{blake2010adrenal}. 
Figure~\ref{fig:intro_pcc_types} illustrates the portal venous phase CT volumes containing PCC tumors from different genetic clusters. A 2D slice from each volume is displayed with a soft tissue window center and width of [50, 450] HU.

Segmentation of pheochromocytoma tumors from abdominal CT scans plays a pivotal role in clinical diagnosis and treatment planning. Precise segmentation facilitates accurate measurement of tumor volume and characterization of morphological features, thereby enhancing clinical decision-making and improving prognostic assessment~\cite{gillies2016radiomics}. Further, accurate segmentation of PCC tumors may also facilitate predicting the genetic cluster associated with the tumor, enabling low-cost identification as opposed to an expensive genetic testing.  Some recent approaches demonstrated performance on genetic type classification of PCC tumors from the manually segmented tumor regions~\cite{santra2023anatomical,makroo2025enhanced,zhou2025ct}. However, manual segmentation of pheochromocytomas is labor-intensive, time-consuming, and subject to inter-observer variability, particularly given the often heterogeneous and irregular tumor boundaries observed in clinical practice~\cite{menze2014multimodal}. Thus, automated segmentation methods using advanced computational techniques are increasingly important to overcome these challenges, enhance reproducibility, and streamline clinical workflows~\cite{litjens2017survey}.

Recent advances in artificial intelligence (AI), particularly deep learning-based segmentation approaches such as U-Net and its variants, have demonstrated substantial improvements in medical image segmentation tasks, including tumor delineation~\cite{ronneberger2015u,hatamizadeh2022unetr, hatamizadeh2021swin}. Despite these advancements, the segmentation of pheochromocytomas remains challenging due to variability in tumor size, shape, contrast uptake patterns, and proximity to adjacent anatomical structures, making it difficult for traditional and even some contemporary automated methods to achieve consistently high accuracy. Among recent AI-based segmentation methods, the nnU-Net segmentation method is the most widely adopted in various biomedical segmentation applications because of its high accuracy and efficiency~\cite{isensee2021nnu}. While the literature on pheochromocytoma segmentation from abdominal CT is limited, some prior approaches proposed training nnU-Net exploiting 2D or pseudo-3D annotations~\cite{oluigbo2024weakly,oluigbo2024weaklyarxiv}. Additionally, those approaches addressed PCC detection by predicting bounding boxes, rather than precise tumor delineation or segmentation. Such detection-based 2D/pseudo-3D models are not suitable for the accurate quantification of key tumor characteristics, such as volume, boundary irregularity, and changes in standardized uptake values (SUV), which are crucial for assessing tumor progression. 

In addition, to enhance the segmentation performance of a U-Net–based network for a specific task, some approaches propose training with multi-class labels, and then extracting only the target class prediction after inference. The auxiliary labels in such a labeling strategy are considered anatomical priors, for example, nearby organs around the target lesions to be segmented, or surrounding fat and muscle tissues in subcutaneous edema segmentation~\cite{bhadra2024subcutaneous, mathai2024segmentation}. Such prior information can provide the context of regions around the target object, and may also reduce the potential false positives in the segmentation mask. However, previous detection-based approaches on PCC segmentation~\cite{oluigbo2024weakly,oluigbo2024weaklyarxiv} exploited only the body-region annotation around the tumor as an anatomical prior. The effect of surrounding organs as anatomical priors was not studied, which may provide further insight in accurate 3D segmentation of PCC tumors.

In this work, we propose an nnU-Net-based segmentation framework for 3D segmentation of pheochromocytoma tumors from contrast-enhanced abdominal CT scans. We investigate the effect of a wide variety of anatomical prior combinations on improving the segmentation performance in a multiclass setting. We also demonstrate the quantification of PCC tumor burden from the segmentation mask as compared to the reference measures from ground-truth annotations, which can be an essential clinical attribute for analyzing disease progression.


\newpage
\section{Methods}
\label{sec2}

\subsection{Patient Sample}

In this retrospective study, one dataset containing abdominal contrast-enhanced CT (CECT) scans was used. The study was Health Insurance Portability and Accountability Act (HIPAA) compliant and approved by our institution's Institutional Review Board. The requirement for signed informed consent from the patients was waived. The dataset comprised 105 portal venous phase CECT series from 91 patients who underwent imaging at the National Institutes of Health (NIH) Clinical Center between 1999 and 2022. All studies were acquired using a standardized imaging protocol with a fixed 70-second delay following intravenous administration of contrast material. Among the 105 series, 35 were associated with Kinase subtype, 25 with \textit{VHL/EPAS1} subtype, 29 with sporadic subtype, and 16 with \textit{SDHx} subtype, respectively. Across the 105 series, a total of 135 PCC were identified by a senior radiologist. 

\subsection{Generation of Annotation Reference Standard}
\label{subsec1}

\noindent
\textbf{Pheochromocytoma Labeling.} The reference annotation for the dataset was obtained using three experts: one senior board-certified radiologist (30+ years of experience) and two first year radiology residents. First, the senior radiologist manually identified each PCC in the 105 CECT volumes and the 3D center coordinate of each lesion was recorded. Then, the two radiology residents used the recorded 3D positions to manually annotate the full 3D extent of the lesions in 50\% of the dataset (52 volumes). Manual annotation was performed using the open-source software ITK-SNAP~\cite{yushkevich2006user}. Next, an iterative learning framework was leveraged to reduce the cumbersome and time-consuming nature of annotations performed by the residents. Using the 52 volumes, a 3D full-resolution nnU-Net model was trained with a single class (tumor vs. background). The trained model was then run on the remaining CECT volumes to identify PCC. The resulting predictions were manually refined by the residents to correct any segmentation errors and added back to the labeled training data subset. 

\begin{figure}[htbp]
  \centering
  \resizebox{0.95\textwidth}{!}{%
    \begin{minipage}{\textwidth}
      \centering

      \begin{minipage}[b]{0.28\textwidth}
        \centering
        \includegraphics[width=\linewidth]{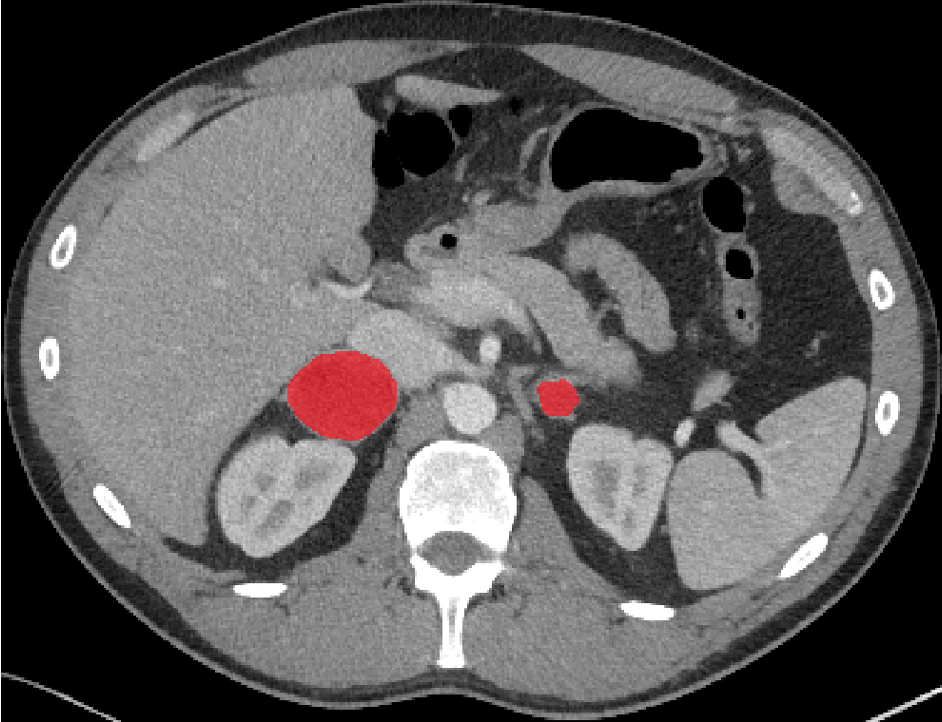}
        \vspace{-1mm}
        (a)
      \end{minipage}
      \hfill
      \begin{minipage}[b]{0.28\textwidth}
        \centering
        \includegraphics[width=\linewidth]{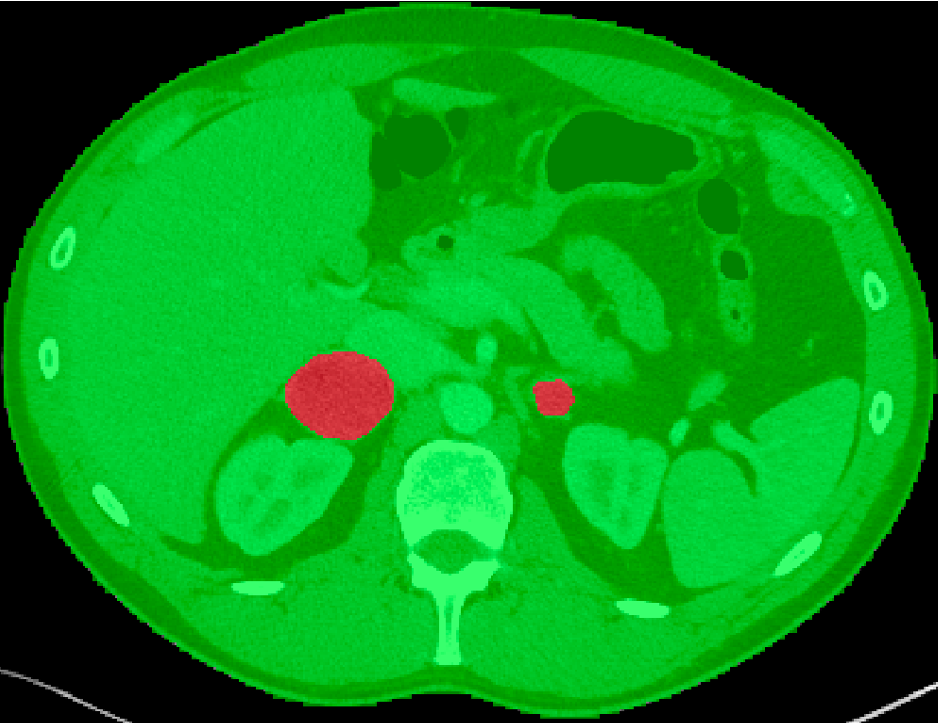}
        \vspace{-1mm}
        (b)
      \end{minipage}
      \hfill
      \begin{minipage}[b]{0.28\textwidth}
        \centering
        \includegraphics[width=\linewidth]{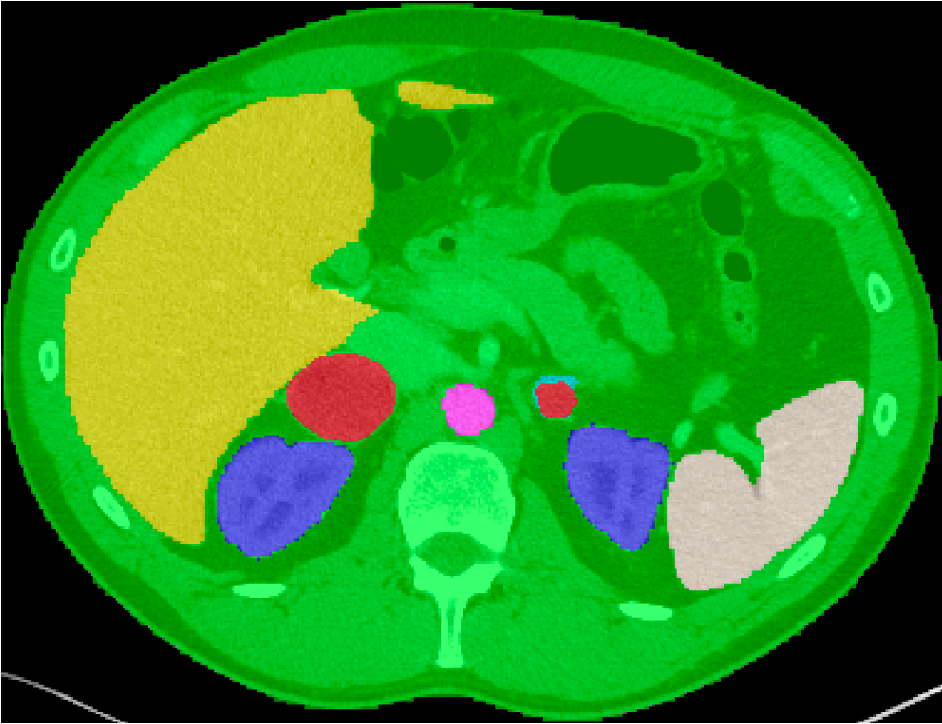}
        \vspace{-1mm}
        (c)
      \end{minipage}

      \vspace{1mm}

      \begin{minipage}[b]{0.28\textwidth}
        \centering
        \includegraphics[width=\linewidth]{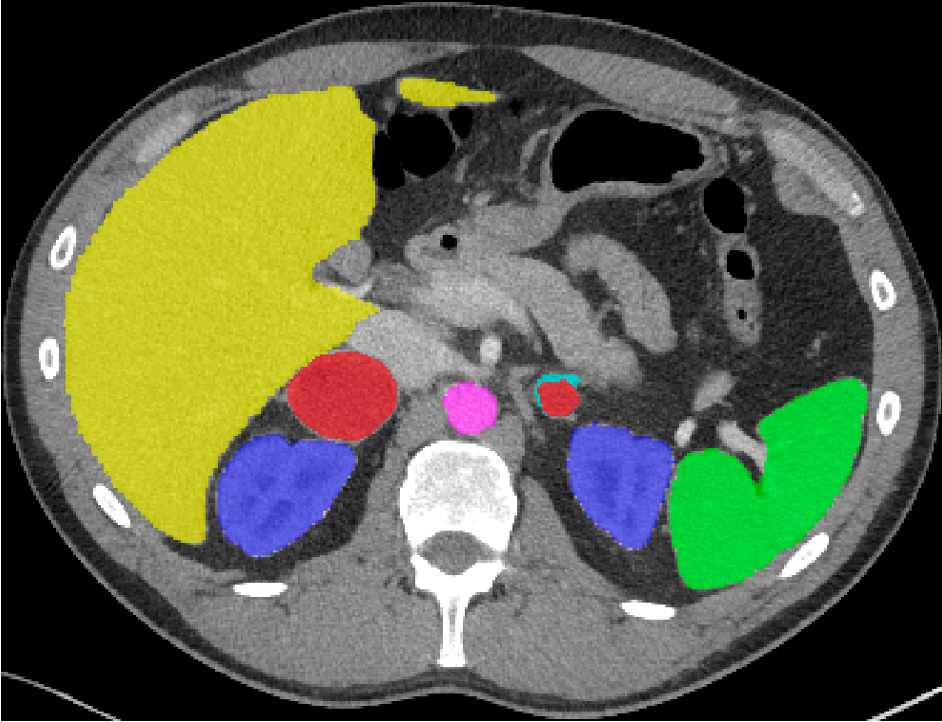}
        \vspace{-1mm}
        (d)
      \end{minipage}
      \hfill
      \begin{minipage}[b]{0.28\textwidth}
        \centering
        \includegraphics[width=\linewidth]{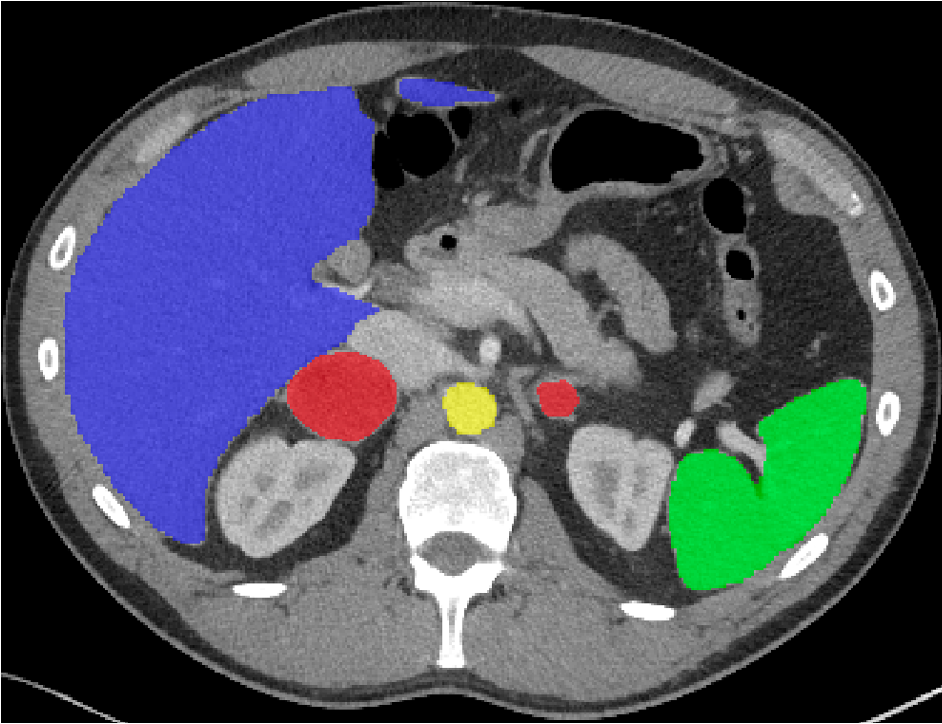}
        \vspace{-1mm}
        (e)
      \end{minipage}
      \hfill
      \begin{minipage}[b]{0.28\textwidth}
        \centering
        \includegraphics[width=\linewidth]{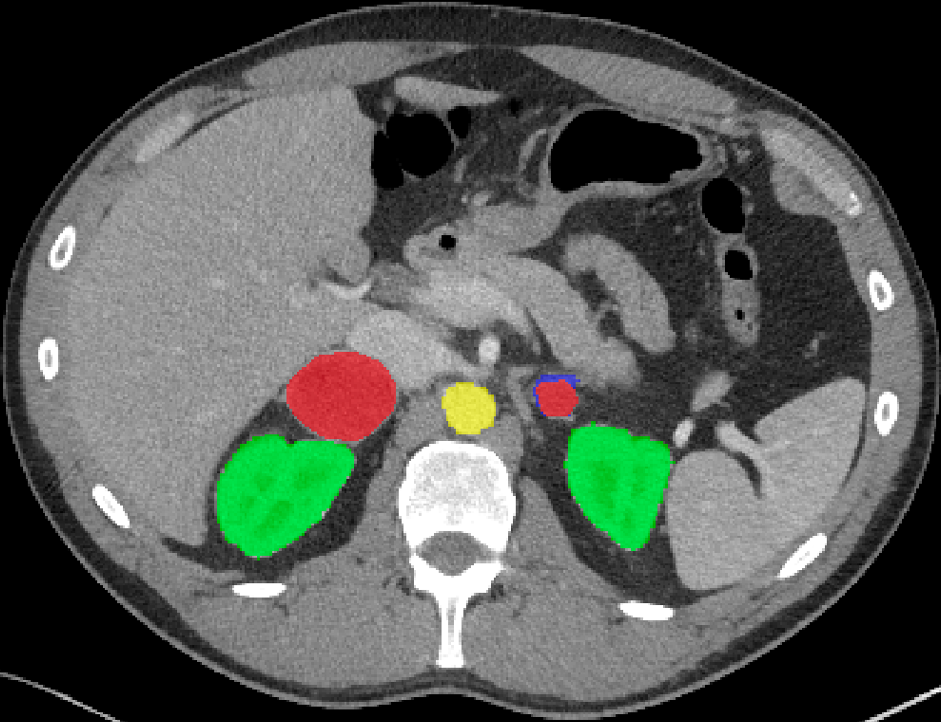}
        \vspace{-1mm}
        (f)
      \end{minipage}

      \vspace{1mm}

      \begin{minipage}[b]{0.28\textwidth}
        \centering
        \includegraphics[width=\linewidth]{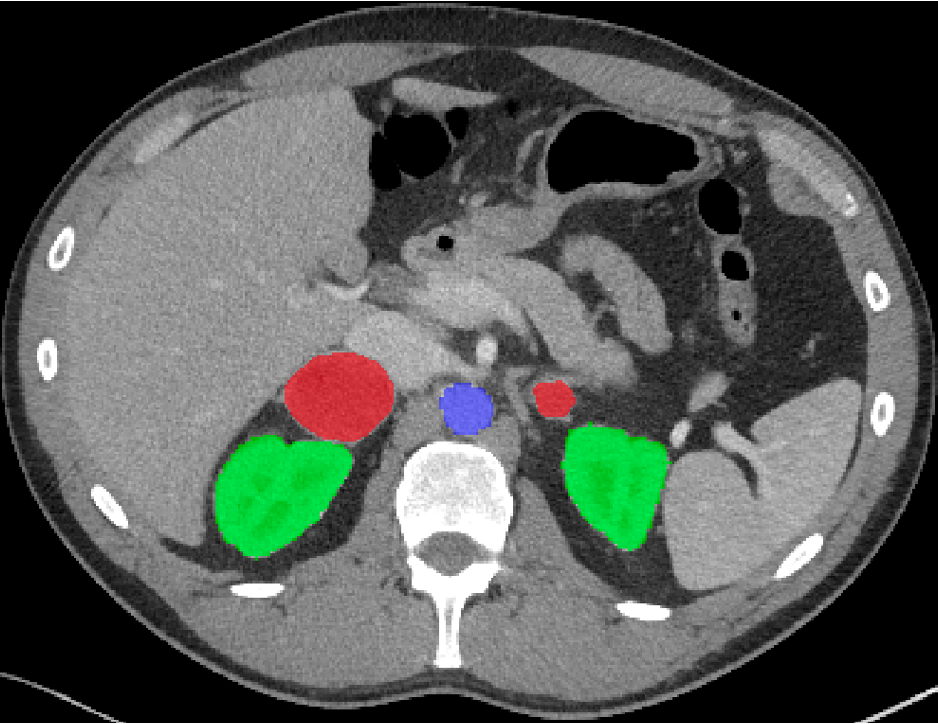}
        \vspace{-1mm}
        (g)
      \end{minipage}
      \hfill
      \begin{minipage}[b]{0.28\textwidth}
        \centering
        \includegraphics[width=\linewidth]{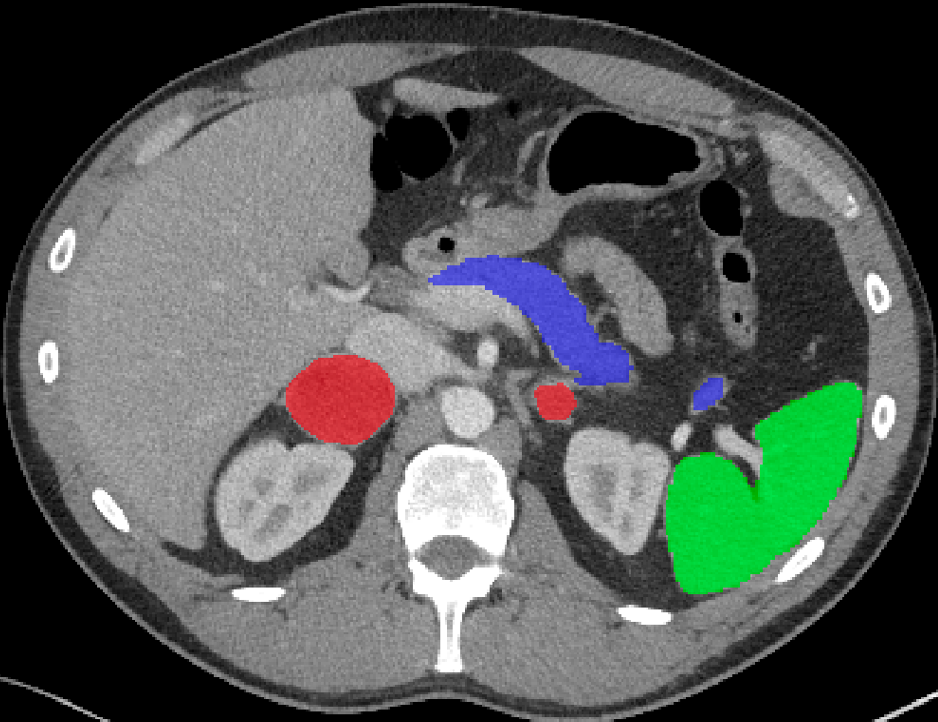}
        \vspace{-1mm}
        (h)
      \end{minipage}
      \hfill
      \begin{minipage}[b]{0.28\textwidth}
        \centering
        \includegraphics[width=\linewidth]{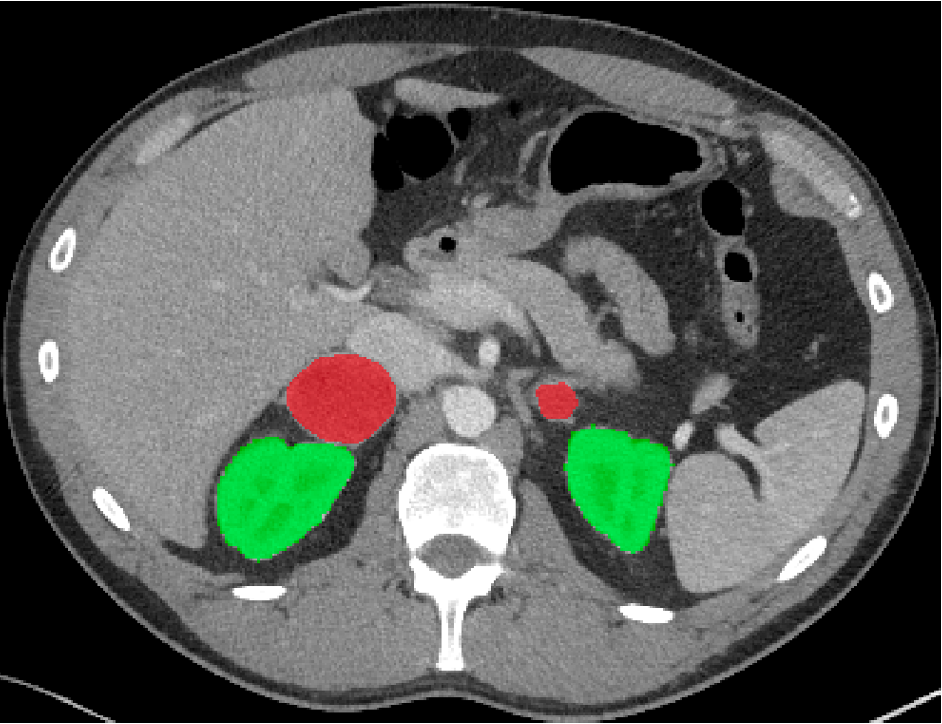}
        \vspace{-1mm}
        (i)
      \end{minipage}

      \vspace{1mm}

      \begin{minipage}[b]{0.28\textwidth}
        \centering
        \includegraphics[width=\linewidth]{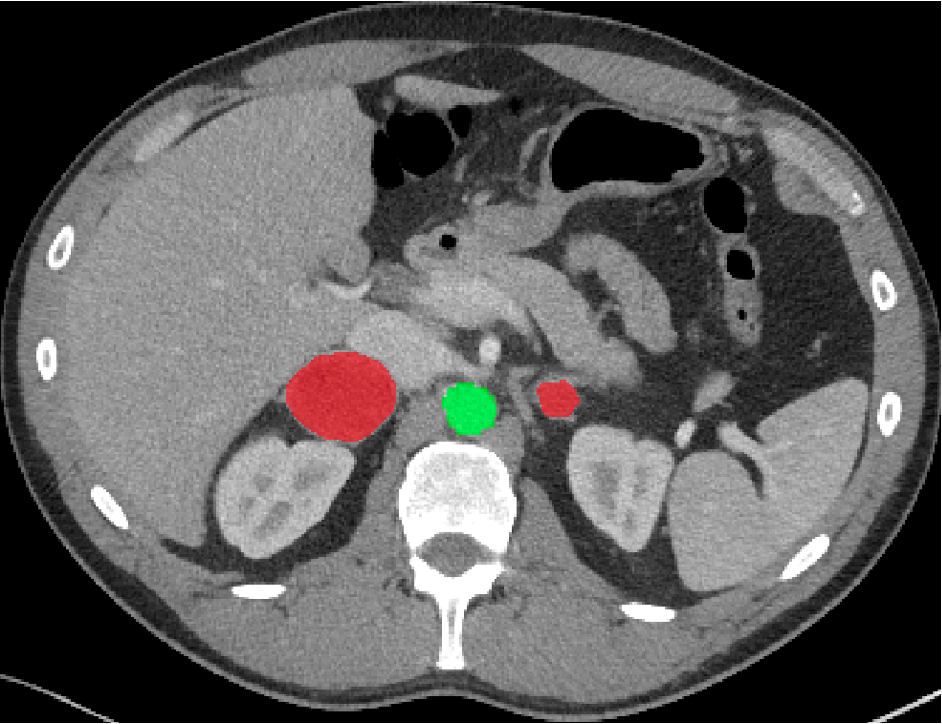}
        \vspace{-1mm}
        (j)
      \end{minipage}
      \hspace{0.06\textwidth}
      \begin{minipage}[b]{0.28\textwidth}
        \centering
        \includegraphics[width=\linewidth]{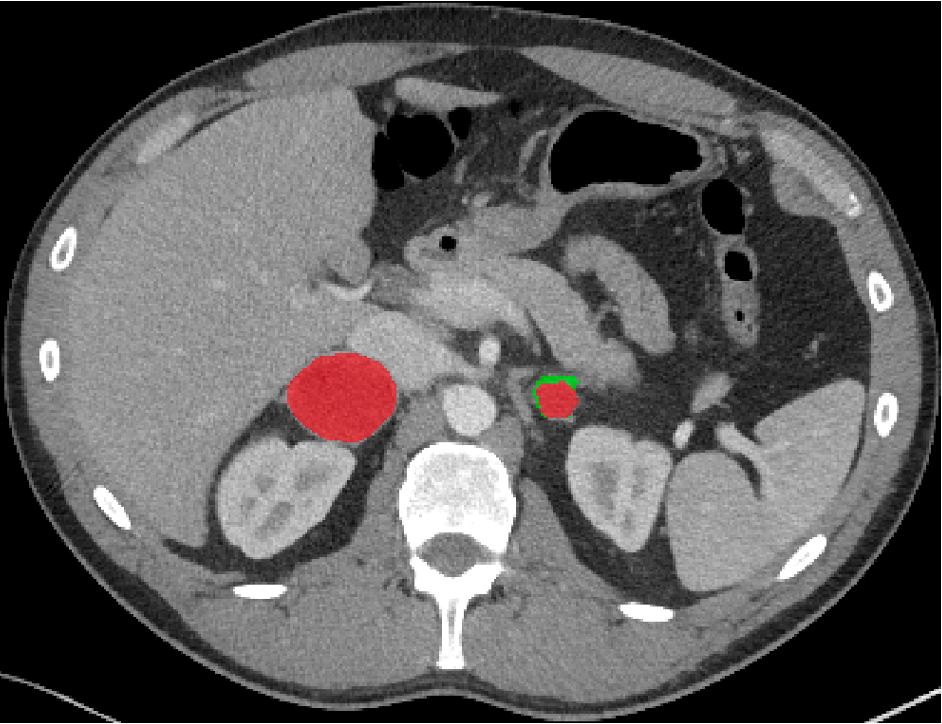}
        \vspace{-1mm}
        (k)
      \end{minipage}

    \end{minipage}
  }
  \caption{
    Eleven types of annotations studied: 
    (a) Tumor only [T], 
    (b) Tumor + Body [TB], 
    (c) Tumor + Body + Liver + Spleen + Kidney + Aorta + Adrenal Gland [TBLSKAG], 
    (d) Tumor + Liver + Spleen + Kidney + Aorta + Adrenal Gland [TLSKAG], 
    (e) Tumor + Liver + Spleen + Aorta [TLSA], 
    (f) Tumor + Kidney + Aorta + Adrenal Gland [TKAG], 
    (g) Tumor + Kidney + Aorta [TKA], 
    (h) Tumor + Spleen + Pancreas [TSP], 
    (i) Tumor + Kidney [TK], 
    (j) Tumor + Aorta [TA], 
    (k) Tumor + Adrenal Gland [TG].
    Note that the same organ may appear in different colors across subfigures because the number of classes varies across annotation strategies, and colors were automatically assigned based on the class index order within each strategy.
  }
  \label{fig:method_annotation_type}
\end{figure}

\medskip
\noindent
\textbf{Anatomy Priors.} The publicly available TotalSegmentator tool (henceforth called TS) \cite{wasserthal2023totalsegmentator} was used as it can robustly segment 117 different anatomical structures (``total'' task) and the body trunk (``body'' task) on CT. Eleven different training label combinations were created by merging the lesions with the abdominal organs and the body region masks. These combinations are the following: (1) Tumor only [T], (2) Tumor + Body [TB], (3) Tumor + Body + Liver + Spleen + Kidney + Aorta + Adrenal Gland [TBLSKAG], (4) Tumor + Liver + Spleen + Kidney + Aorta + Adrenal Gland [TLSKAG], (5) Tumor + Liver + Spleen + Aorta [TLSA], (6) Tumor + Kidney + Aorta + Adrenal Gland [TKAG], (7) Tumor + Kidney + Aorta [TKA], (8) Tumor + Spleen + Pancreas [TSP], (9) Tumor + Kidney [TK], (10) Tumor + Aorta [TA], and (11) Tumor + Adrenal Gland [TG]. Figure~\ref{fig:method_annotation_type} shows the different training labels that were studied. The rationale behind the use of anatomy priors was to study the effect of: (1) tumor labels only versus multi-class labels, and (2) performance of body-only anatomical prior versus additional anatomical priors of organs. 

\subsection{Semantic Segmentation with 3D nnU-Net}\label{subsec2}

\noindent
\textbf{Model.} The publicly available nnU-Net framework~\cite{isensee2021nnu} was employed to segment pheochromocytomas on abdominal CECT. The dataset pre-processing stage involves adaptive handling of the training CT volumes, guided by a fingerprinting strategy that captures key dataset-specific characteristics. This process analyzed voxel spacing, intensity distributions, and image dimensions to determine the optimal resampling resolution and the intensity normalization parameters. Foreground voxel intensities across all training cases were used to compute the 0.5 and 99.5 percentiles, as well as the mean and standard deviation. These values were then used for percentile clipping and z-score normalization. Additionally, based on the dataset fingerprint, nnU-Net automatically configured the patch size, batch size, and network depth to ensure adequate receptive field coverage and stable gradient estimation during training.

After dataset pre-processing, 3D full-resolution nnU-Net models were trained to segment PCC using the aforementioned training label combinations. The network architecture was automatically configured based on the dataset properties. During training, the model learned to estimate the semantic (or voxel-wise) segmentation masks given the input CT volumes by minimizing the error between the predictions and the corresponding ground-truth annotations. Optimization was carried out using a composite loss function $\mathcal{L}$, combining cross-entropy and soft Dice loss (Eq. ~\ref{eq:total_loss}). The cross-entropy component promoted consistent learning across all voxels, but may be biased toward dominant classes in imbalanced settings. The Dice loss counteracted the class imbalance by directly maximizing the spatial overlap between predicted ($\hat{y}$) and ground-truth ($y$) regions, placing stronger emphasis on underrepresented structures like PCCs. In Equations~\ref{eq:cross_entropy}--\ref{eq:dice_loss}, $N$ is the total number of voxels, $C$ is the number of classes, $y_{i,c}$ refers to ground truth (1 if voxel $i$ belongs to class $c$, else 0), and $\hat{y}_{i,c}$ denotes the predicted probability of voxel $i$ for class $c$. 

\begin{equation}
\mathcal{L} = \mathcal{L}_{\text{CE}} + \mathcal{L}_{\text{Dice}}
\label{eq:total_loss}
\end{equation}

\begin{equation}
\mathcal{L}_{\text{CE}} = -\frac{1}{N} \sum_{i=1}^{N} \sum_{c=1}^{C} y_{i,c} \log(\hat{y}_{i,c})
\label{eq:cross_entropy}
\end{equation}

\begin{equation}
\mathcal{L}_{\text{Dice}} = 1 - \frac{1}{C} \sum_{c=1}^{C} \frac{2 \sum_{i=1}^{N} y_{i,c} \hat{y}_{i,c}}{\sum_{i=1}^{N} y_{i,c} + \sum_{i=1}^{N} \hat{y}_{i,c}}
\label{eq:dice_loss}
\vspace{1em}
\end{equation}

During inference, the trained nnU-Net model was applied to unseen abdominal CT volumes to segment PCC. Only the PCC labels from the predicted segmentation masks were considered, while other labels were discarded.

\subsection{Experimental Settings}

\noindent
\textbf{Dataset Division.} The dataset was partitioned into a 70\%-30\% split, with 74 scans allocated for training (Kinase: 26, \textit{VHL/EPAS1}: 20, Sporadic: 18, \textit{SDHx}: 10) and 31 scans reserved for testing (Kinase: 8, \textit{VHL/EPAS1}: 5, Sporadic: 11, \textit{SDHx}: 6). The division was performed randomly at the patient level, ensuring no patient data leakage. Scans from the same patient were assigned to either the training or test data subset.  

\medskip
\noindent
\textbf{Training Parameters.} Training was performed using the nnU-Net v2.5.2 framework with a 3D full-resolution configuration. The model was trained with a batch size of 2 and a patch size of 80$\times$ 160$\times$ 160, using CT volumes resampled to a uniform spacing of [2.5, 0.774, 0.774] mm and normalized with default CT intensity normalization. The architecture followed a 6-stage 3D U-Net (PlainConvUNet) with increasing feature maps from 32 to 320, instance normalization, and LeakyReLU activation. Optimization was performed using stochastic gradient descent (SGD) with an initial learning rate of 0.01, momentum of 0.99, and weight decay of 3 $\times 10^{-5}$. A polynomial learning rate scheduler was used to gradually reduce the learning rate across 1000 training epochs. The model was trained using an NVIDIA A100 GPU in the NIH High Performance Computing (HPC) cluster. 

\medskip
\noindent
\textbf{Comparisons.} The best-performing model from the eleven approaches was first determined. Then, this best model was directly compared against the TB model used in prior work \cite{oluigbo2024weakly}. For this comparison, both models were trained on the entire dataset (105 scans) using five-fold cross-validation. Other deep learning architectures were also evaluated, such as: (1) UNETR (a transformer-based model) ~\cite{hatamizadeh2022unetr} and (2) Swin UNET ~\cite{hatamizadeh2021swin}. These models were implemented with the MONAI framework~\cite{cardoso2022monai}. 

\subsection{Statistical Analysis}

\noindent
\textbf{Detection metrics.} Detection performance was assessed using precision (positive predictive value), recall (sensitivity), and F1 score. Connected-component analysis was applied to individually identify tumor instances in both the predicted segmentations (\(S\)) and ground truth masks (\(G\)). F1 score was computed at various intersection-over-union (IoU) thresholds \cite{bilic2023liver,toma2024deepseeded}:
\[
F_1 = \frac{2 \times \text{Precision} \times \text{Recall}}{\text{Precision} + \text{Recall}}, \quad \text{IoU} = \frac{|\; S \cap G \;|}{|\; S \cup G \;|}
\]

where \(\text{Precision} = \frac{TP}{TP + FP}\), \(\text{Recall} = \frac{TP}{TP + FN}\). Here, \(TP\) denotes correctly detected tumors (IoU above threshold), \(FP\) indicates false positive instances without a matching ground truth, and \(FN\) refers to missed ground truth tumors.

\medskip
\noindent
\textbf{Segmentation metrics.} Segmentation metrics included the Dice Similarity Coefficient (DSC) and Normalized Surface Distance (NSD)~\cite{antonelli2022medical}. DSC measures volumetric overlap between prediction and ground truth, while NSD captures boundary alignment by quantifying the average surface distance within a predefined tolerance. We computed NSD using the standard tolerance value of 1~$mm$. We also computed \textit{p}-values for DSC and NSD using the Wilcoxon signed-rank test, a non-parametric test suitable for paired, non-normally distributed data. Further, PCC tumor burden was determined for each test scan, defined as the total tumor volume in \(\text{cm}^3\) or \(\text{mm}^3\)~\cite{chen2023mri,fleckenstein20163d}. It was quantified by multiplying the number of tumor-specific voxels (\(N_{\text{tumor}}\)) with the voxel volume (\(s_x \times s_y \times s_z\)), where \(s_x\), \(s_y\), and \(s_z\) denote the voxel spacings in millimeters. 
\[
\text{Tumor Burden}_{\text{cm}^3} = \frac{N_{\text{tumor}} \times s_x \times s_y \times s_z}{1000}
\]

\newpage
\section{Results}\label{sec4}

\subsection{Detection Results} 

Table~\ref{tab:instance_accuracy} summarizes the instance-wise detection performance of the eleven different models (computed using an IoU threshold of 0.5). The TKA model (tumor + kidney + aorta) achieved the highest scores with 0.825 precision, 0.892 sensitivity (recall) and 0.857 F1 score, respectively. The TK model (tumor + kidney) also achieved a high recall (0.892), although at the cost of lower precision (0.750) compared to TKA. In contrast, the TB model (tumor + body) showed the lowest F1 score (0.681), driven by both lower precision (0.588) and recall (0.811). Adding more anatomy priors (e.g., liver and spleen) reduced the detection performance. Across almost all models, recall was consistently higher than precision, which reflects fewer false negatives (FN) and greater false positives (FP) in this PCC detection task. These results highlight the advantage of the anatomical priors used in the TKA model.

\begin{table}[!h]
\caption{Instance-wise detection accuracy of different models on the test set. Best values in each column are shown in bold.}\label{tab:instance_accuracy}
\centering
\begin{tabular}{@{}llll@{}}
\toprule
\textbf{Method} & \textbf{Precision (↑)} & \textbf{Recall (↑)} & \textbf{F1 score (↑)} \\
\midrule
T       & 0.633 & 0.838 & 0.720 \\
TB      & 0.588 & 0.811 & 0.681 \\
TBLSKAG & 0.667 & 0.757 & 0.709 \\
TLSKAG  & 0.775 & 0.838 & 0.805 \\
TLSA    & 0.769 & 0.811 & 0.790 \\
TKAG    & 0.821 & 0.865 & 0.842 \\
TKA     & \textbf{0.825} & \textbf{0.892} & \textbf{0.857} \\
TSP     & 0.769 & 0.811 & 0.790 \\
TK      & 0.750 & \textbf{0.892} & 0.814 \\
TA      & 0.659 & 0.784 & 0.716 \\
TG      & 0.721 & 0.838 & 0.775 \\
\botrule
\end{tabular}
\end{table}

\begin{figure}[!htbp]
    \centering
    \includegraphics[width=0.65\textwidth]{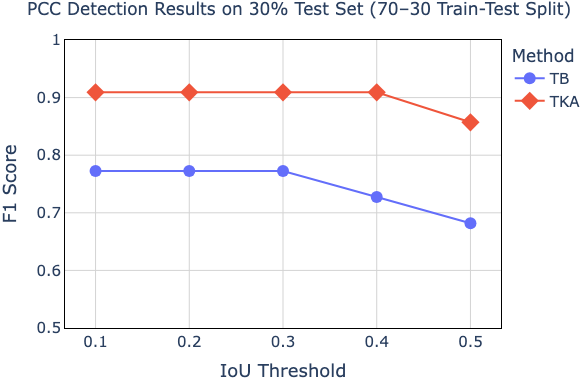}
    \caption{Comparison of detection performances between the TKA and TB methods at different intersection-over-union thresholds on the 30\% test data subset.}
    \label{fig:F1_IoU_comparison}
\end{figure}

Figure~\ref{fig:F1_IoU_comparison} shows the variation in instance-wise F1 score for the TB and TKA models across different IoU thresholds, ranging from 0.1 to 0.5. As expected, the F1 score for both methods decreases with increasing IoU threshold due to the stricter matching criterion for true positive detections. Across all IoU thresholds, the TKA method consistently outperforms TB. Notably, TKA maintains an F1 score above 0.9 up to a 0.4 IoU threshold, reflecting its robustness in detecting tumors even under more stringent spatial overlap conditions. In contrast, the TB model shows a more pronounced decline in F1 score as the IoU threshold increases. 

\begin{figure}[!htbp]
    \centering
    \includegraphics[width=0.65\textwidth]{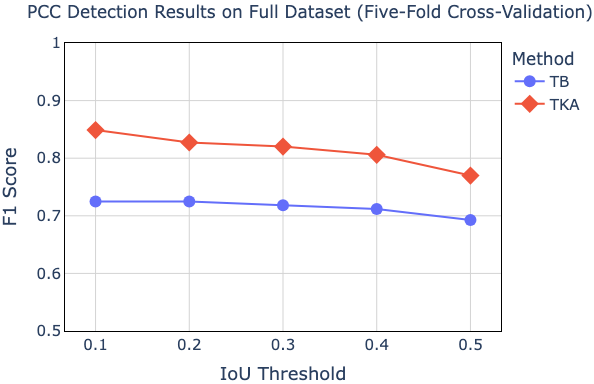}
    \caption{Comparison of detection performances between the TKA and TB methods at different intersection-over-union thresholds on the full dataset of 105 scans.}
    \label{fig:F1_IoU_comparison_fulldataset}
\end{figure}

Finally, Figure~\ref{fig:F1_IoU_comparison_fulldataset} shows instance-wise F1 score comparison across IoU thresholds (0.1–0.5), based on five-fold cross-validation on the full dataset. Similar to the results on the 30\% test data subset, TKA consistently outperforms TB at all thresholds, though the F1 scores are slightly lower due to increased variability across folds. TKA maintained strong performance above (0.8) up to an IoU threshold of 0.4, highlighting its robustness across the full dataset.

\subsection{Segmentation Results} 

\begin{figure}[!htp]
\subfloat[Dice Score Coefficient (DSC)]{%
  \includegraphics[clip,width=\columnwidth]{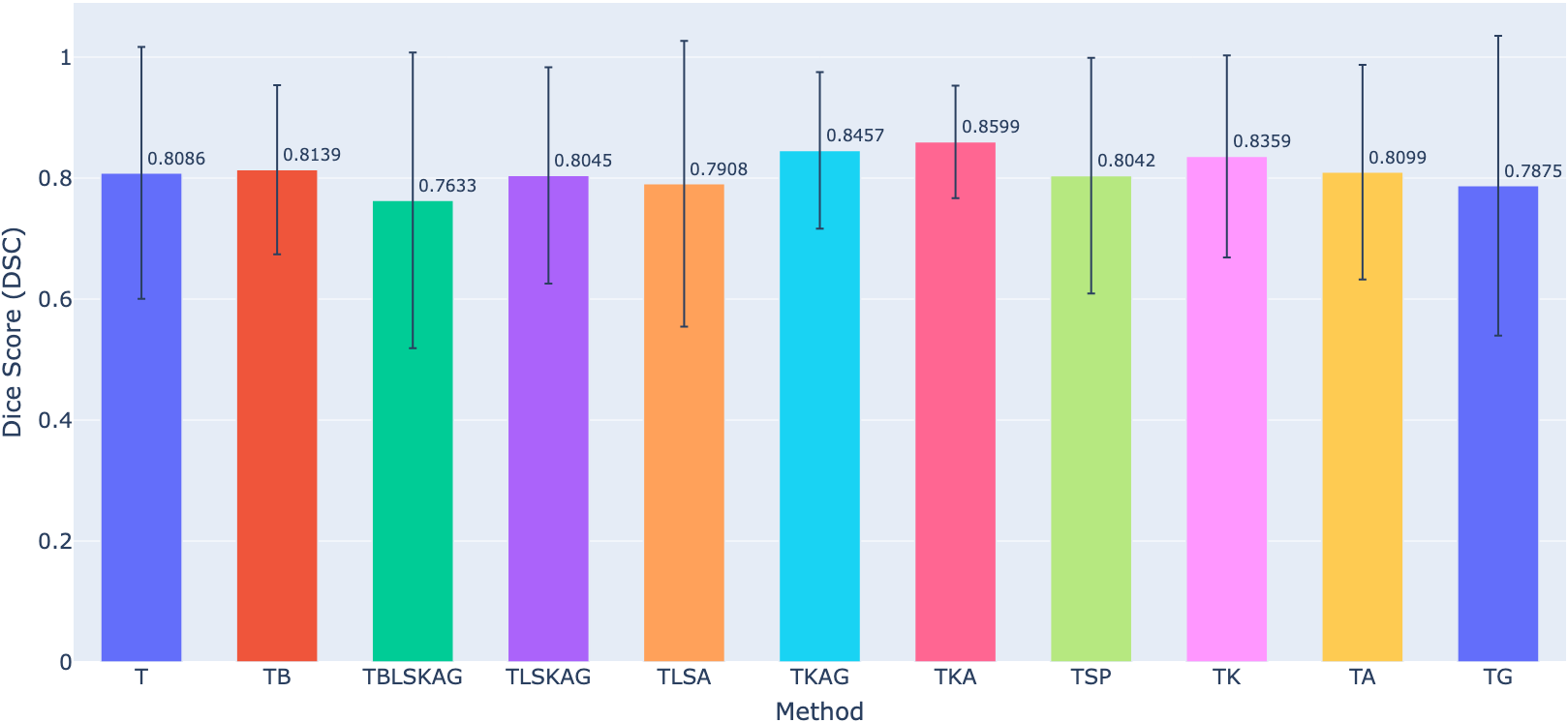}%
}

\subfloat[Normalized Surface Distance (NSD)]{%
  \includegraphics[clip,width=\columnwidth]{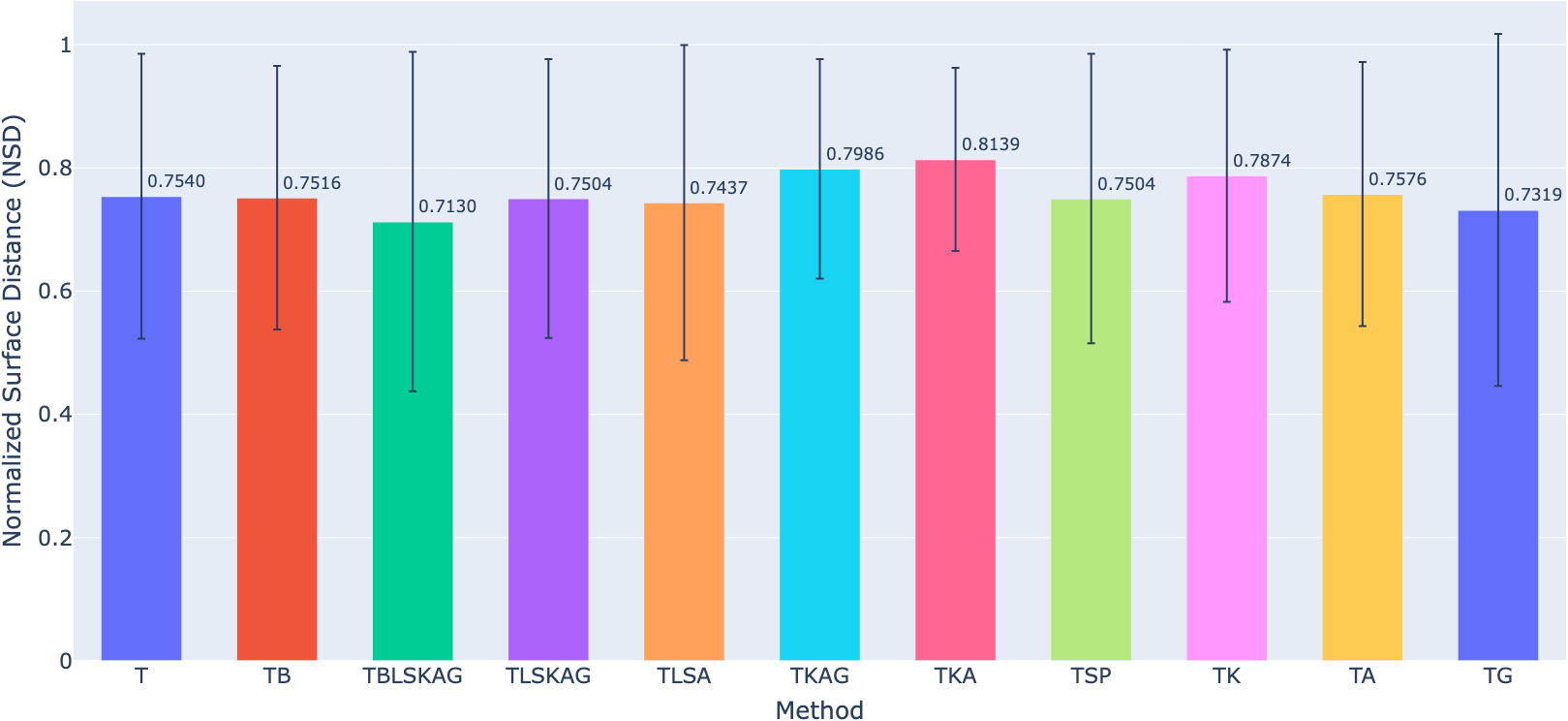}%
}
\caption{Comparison of DSC and NSD (mean and std. dev.) across different models trained with various combinations of training labels.}
\label{fig:dsc_nsd_comparison}
\end{figure}

Figure \ref{fig:dsc_nsd_comparison} shows the DSC and NSD results on the test data subset. 
The TKA model achieved the highest DSC (0.8599) with the lowest standard deviation. Consistent with the DSC results, it also achieved the highest average NSD value (0.8139), reflecting superior boundary alignment between the prediction and the ground truth. It outperformed a prior work's TB model \cite{oluigbo2024weakly}, which relies solely on a broad body region prior. Additionally, the TK and TKAG (with adrenal glands) models also demonstrated strong performance across both DSC and NSD. In contrast, models trained with a large number of classes (e.g., TBLSKAG and TLSKAG), especially those that included large organs like the liver or spleen, showed diminished performance. Furthermore, the TG model, which included only the adrenal glands as anatomical priors, resulted in relatively poor performance.

\begin{figure}[!htbp]
    \centering

    \includegraphics[width=0.24\textwidth]{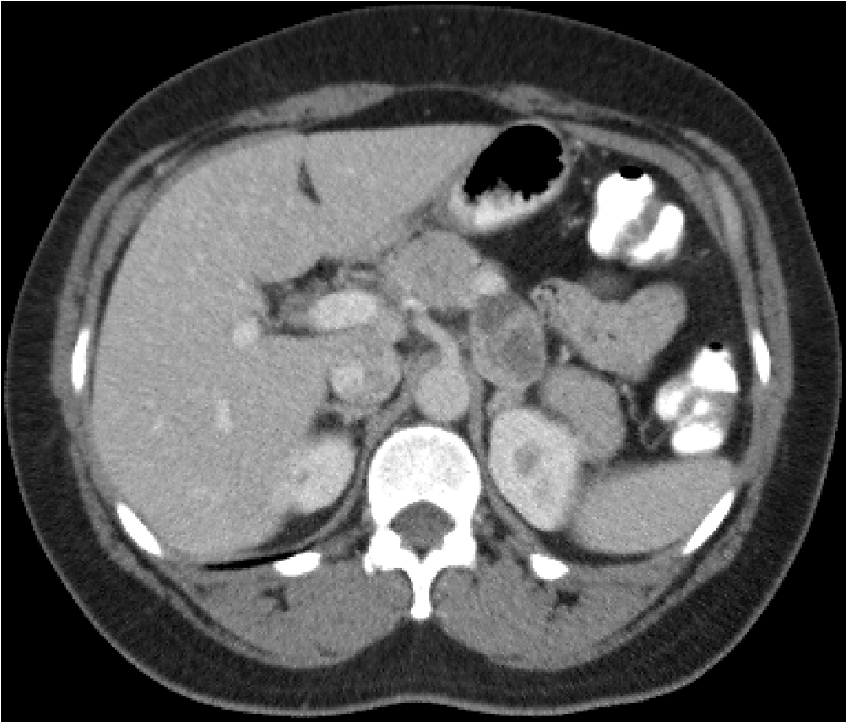}
    \includegraphics[width=0.24\textwidth]{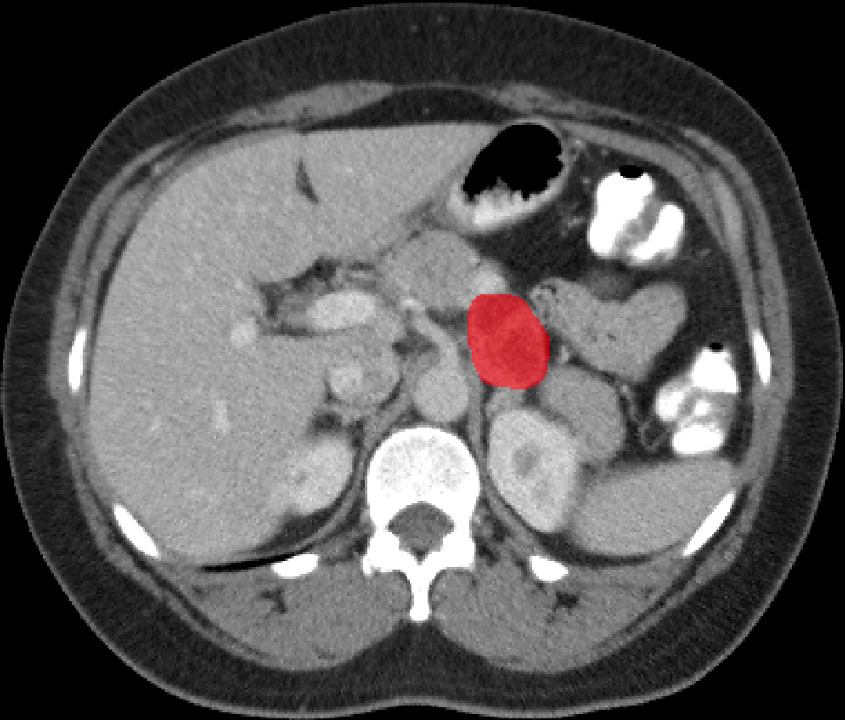}
    \includegraphics[width=0.24\textwidth]{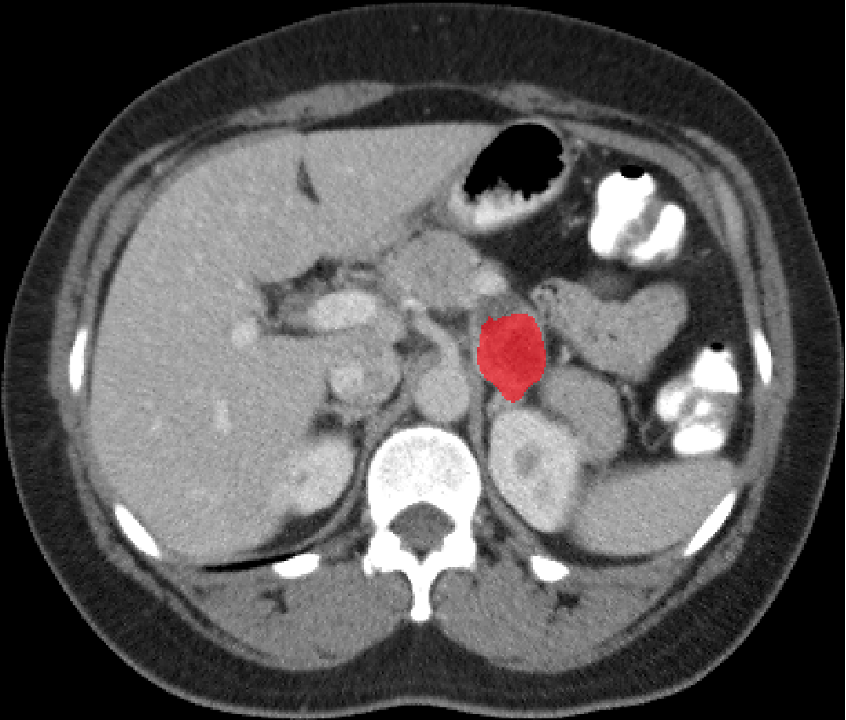}
    \includegraphics[width=0.24\textwidth]{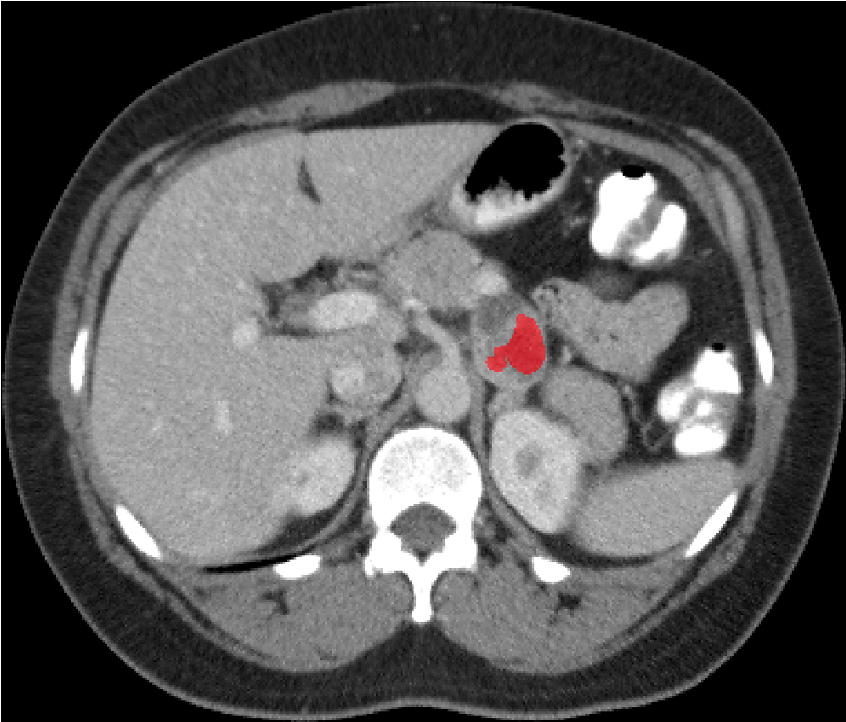} \\[1pt]

    \includegraphics[width=0.24\textwidth]{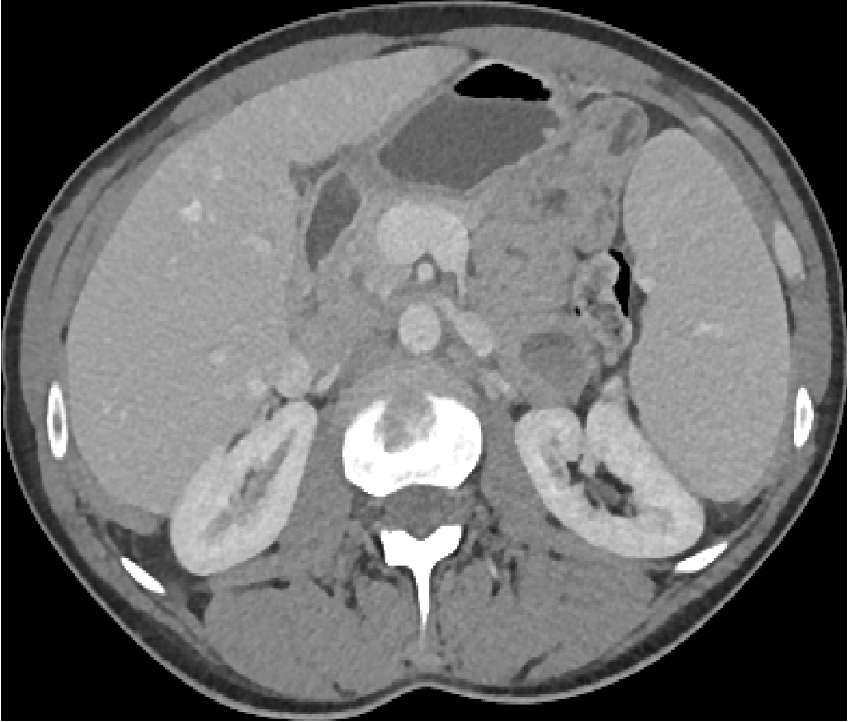}
    \includegraphics[width=0.24\textwidth]{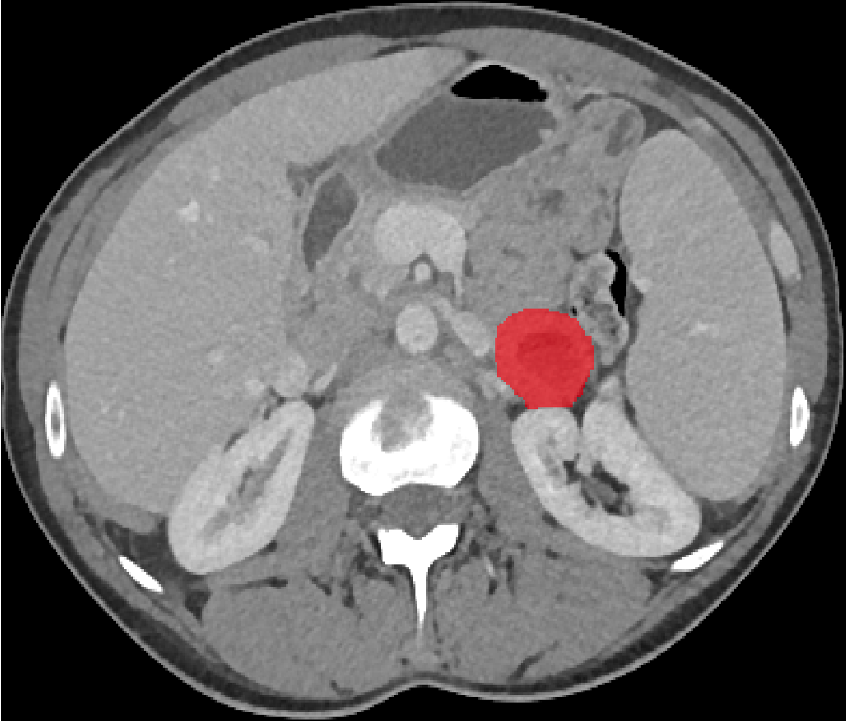}
    \includegraphics[width=0.24\textwidth]{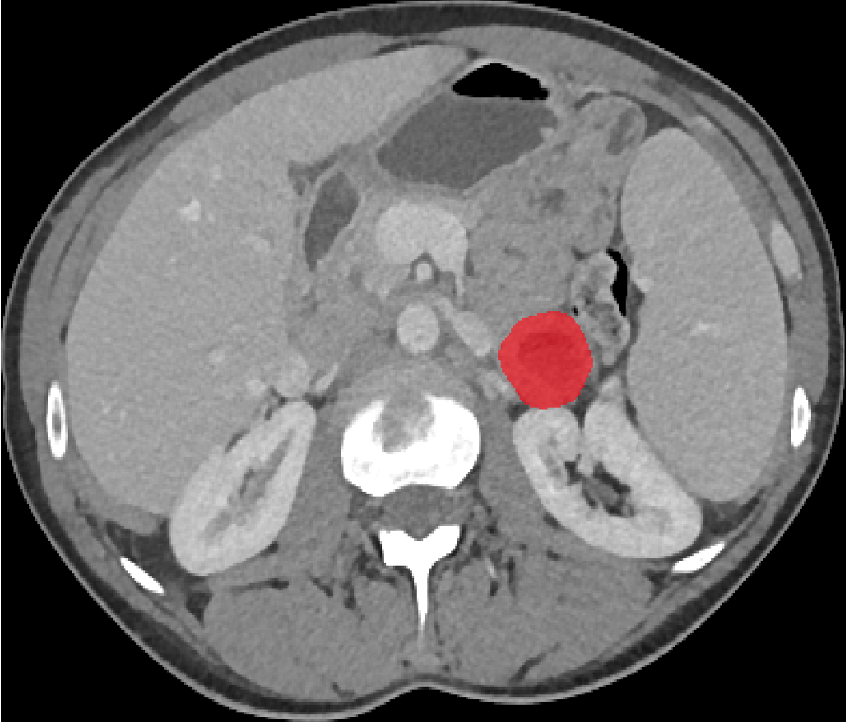}
    \includegraphics[width=0.24\textwidth]{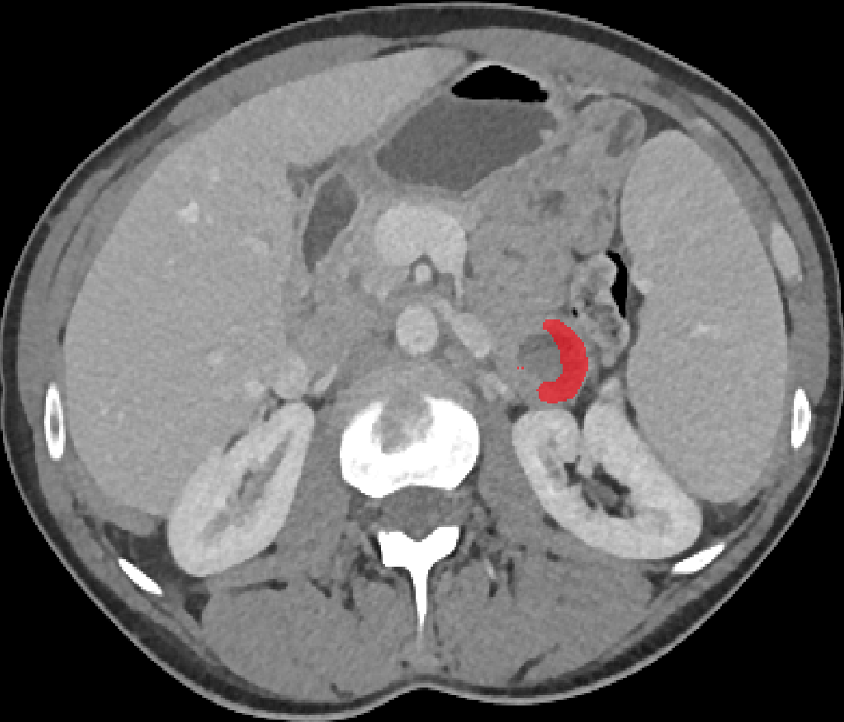} \\[1pt]

    \includegraphics[width=0.24\textwidth]{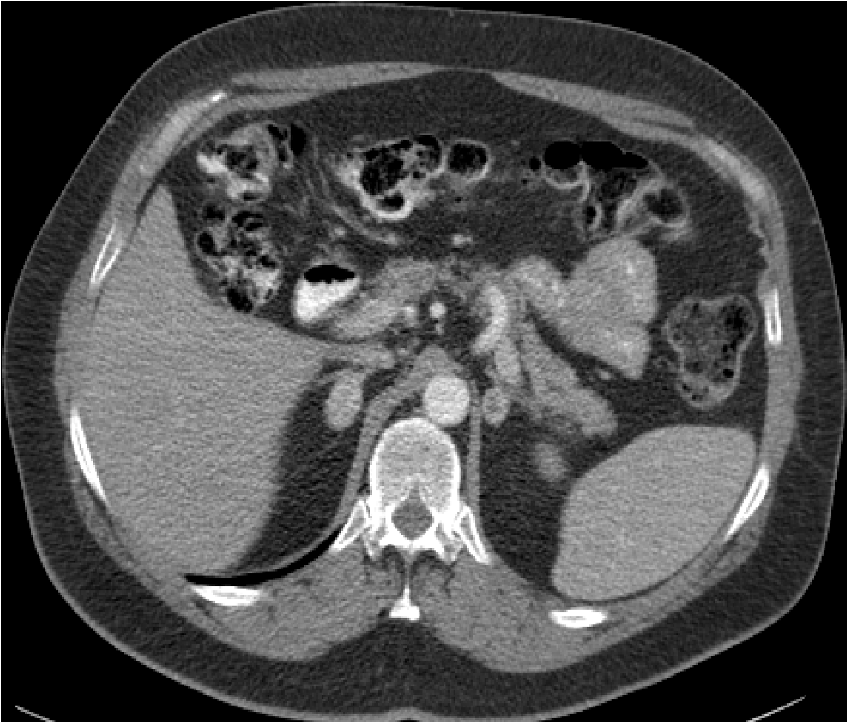}
    \includegraphics[width=0.24\textwidth]{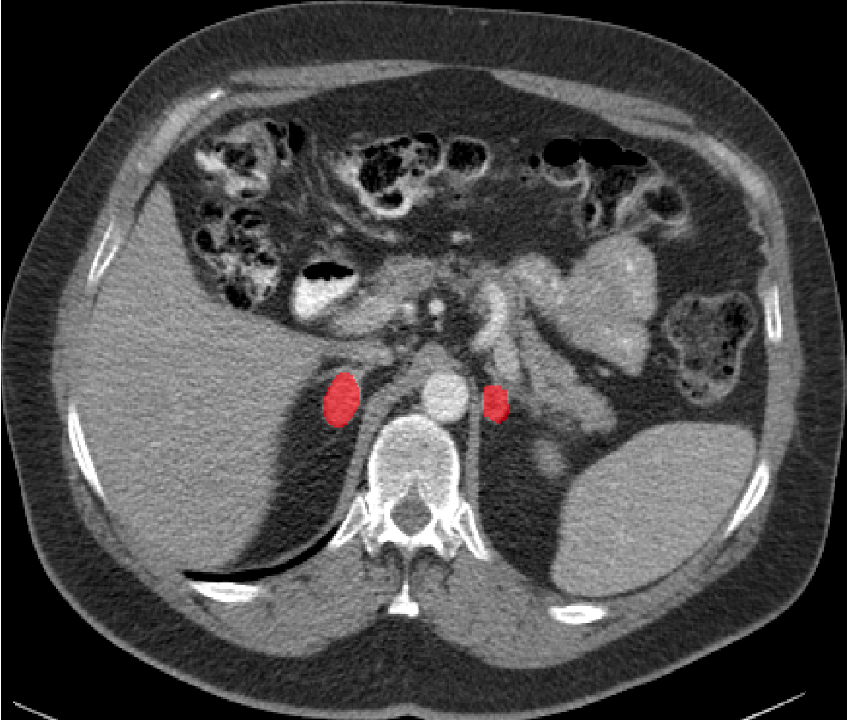}
    \includegraphics[width=0.24\textwidth]{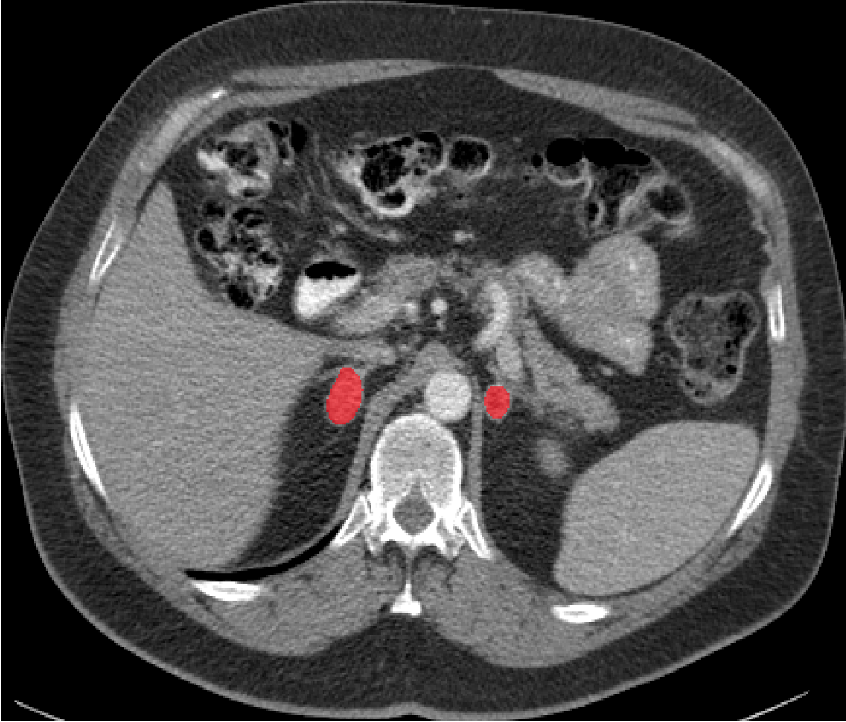}
    \includegraphics[width=0.24\textwidth]{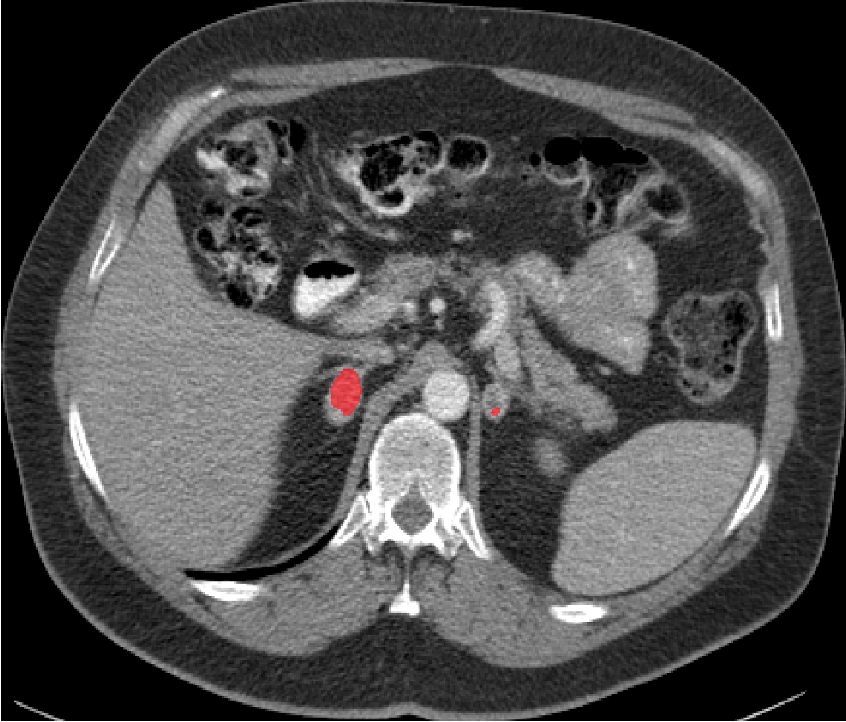} \\[4pt]

    \parbox[t]{0.24\textwidth}{\centering (a) CT Input}
    \parbox[t]{0.24\textwidth}{\centering (b) Ground Truth}
    \parbox[t]{0.24\textwidth}{\centering (c) TKA }
    \parbox[t]{0.24\textwidth}{\centering (d) TB }

    \caption{Qualitative comparison of PCC segmentation across three patients in each row. From left to right: A 2D slice from a CECT scan, corresponding ground truth annotation, and the predictions from two segmentation models (TKA and TB). Note that other anatomy priors are not shown in this figure. }
    \label{fig:tumor_qualitative_comparison}
\end{figure}

Figure~\ref{fig:tumor_qualitative_comparison} shows qualitative PCC segmentation results on representative test cases, comparing TKA and TB predictions against the ground truth. TKA demonstrates superior boundary alignment and more complete tumor coverage, while TB often under-segments the target region.



\begin{figure}[!htbp]
    \centering

    \begin{minipage}[b]{0.65\textwidth}
        \centering
        \includegraphics[width=\textwidth]{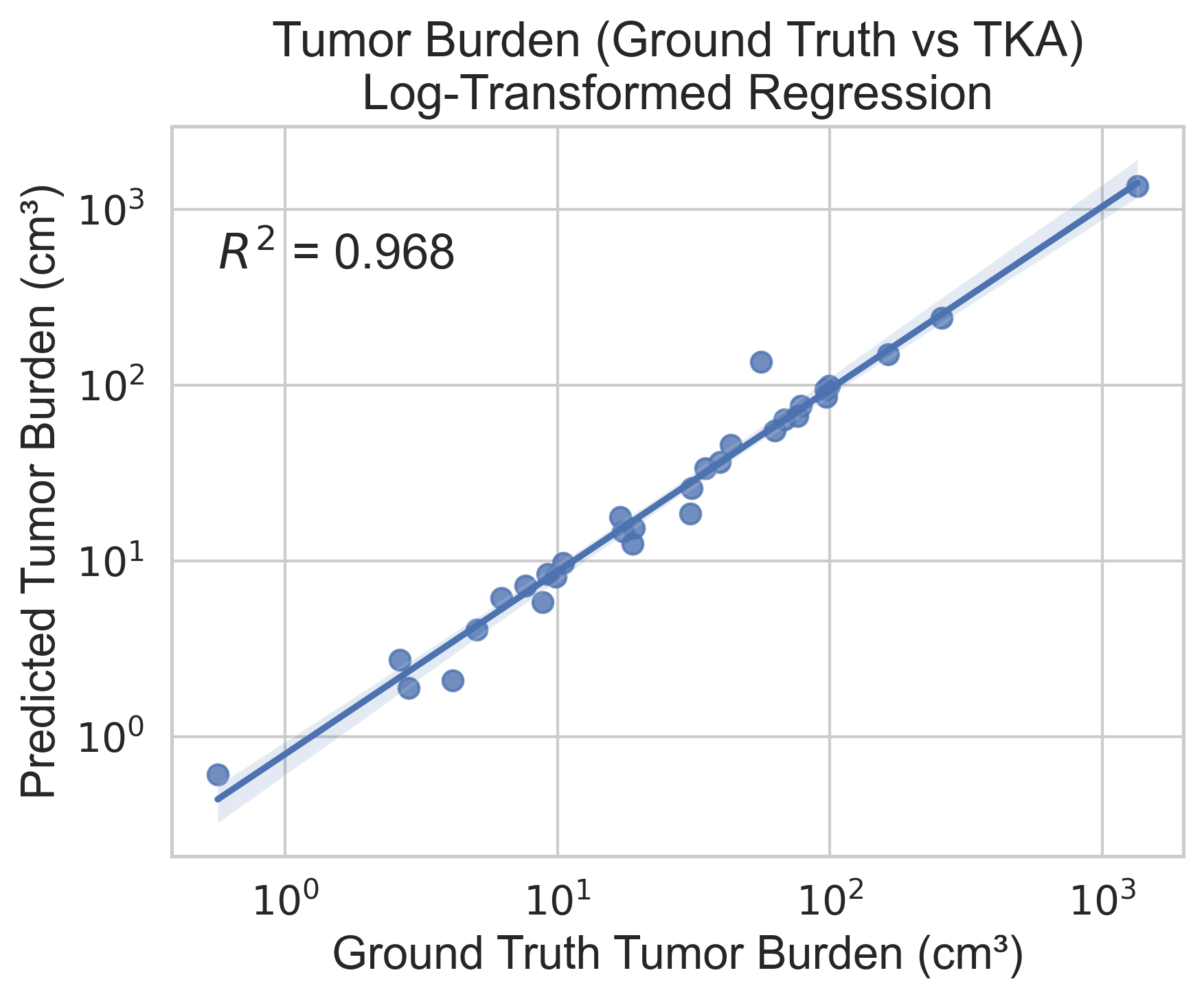}
        \par (a) 
        \label{fig:tka}
    \end{minipage}
    \hfill
    \begin{minipage}[b]{0.65\textwidth}
        \centering
        \includegraphics[width=\textwidth]{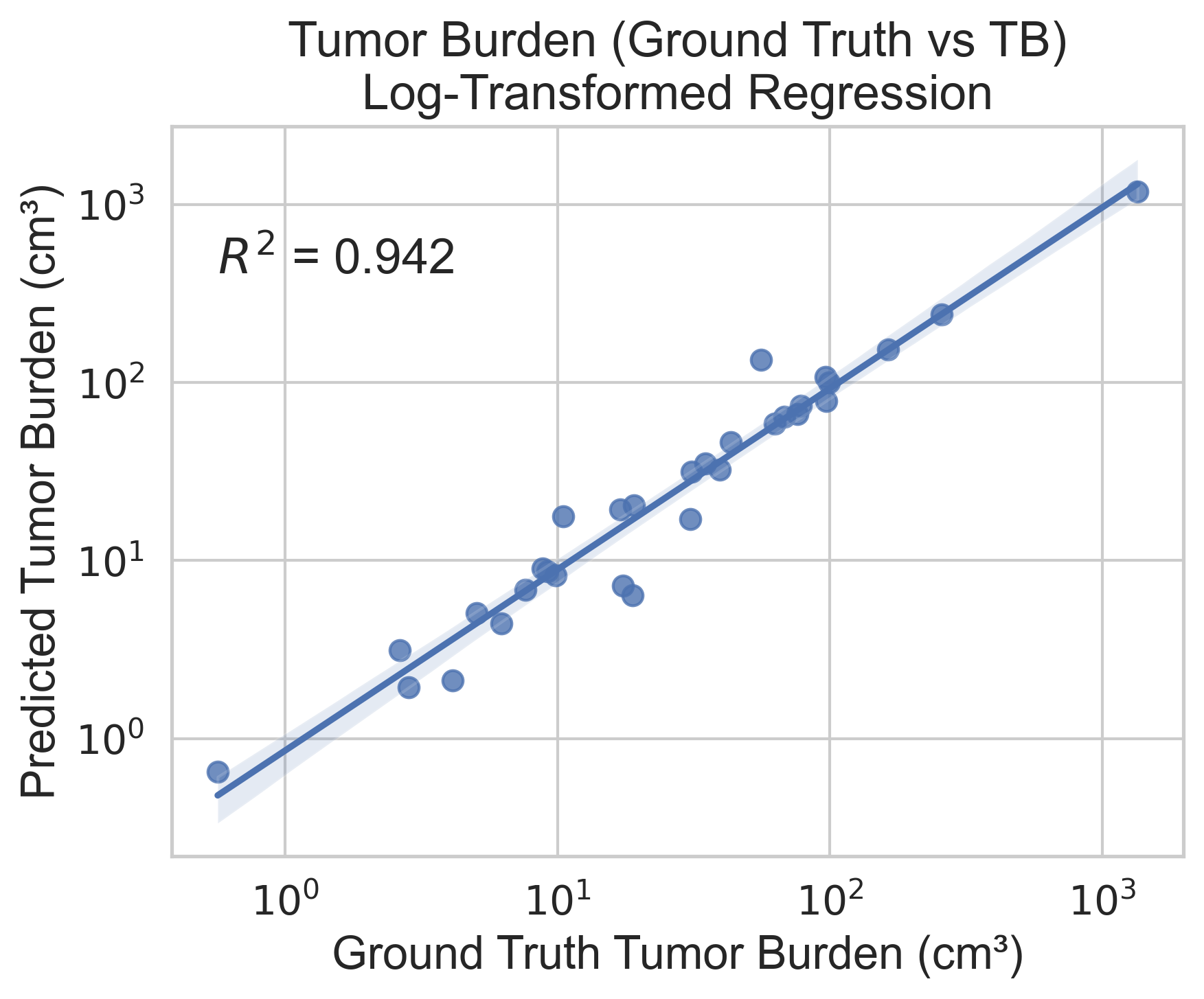}
        \par (b) 
        \label{fig:tb}
    \end{minipage}

    \caption{Scatter plots showing the relationship between ground truth and predicted tumor burden on the test set using annotation strategies: (a) TKA and (b) TB.}
    \label{fig:tka_tb_comparison}
\end{figure}

Figure~\ref{fig:tka_tb_comparison} shows scatter plots comparing predicted tumor burden against the ground truth burden for the TKA and TB models on the test set. Both plots use log-transformed tumor burden values to accommodate the wide range of tumor volumes. The TKA method (Figure~\ref{fig:tka_tb_comparison}a) achieved a higher coefficient of determination (\( R^2 = 0.968 \)), indicating stronger correlation between predicted and ground truth tumor burden compared to the TB method (\( R^2 = 0.942 \); Figure~\ref{fig:tka_tb_comparison}b), which shows greater scatter and larger deviations from the line of perfect agreement, especially for smaller PCC volumes. 

\subsection{Results compared to prior works}

In Table~\ref{tab:network_comparison}, we present the performance comparison of different network architectures for PCC segmentation, evaluated using the TKA annotation strategy. The F1 score was computed at IoU=0.5. Among the three architectures, the nnU-Net achieved the highest performance across all three metrics. Both transformer-based models, UNETR and Swin UNETR, showed comparatively lower performance, with Swin UNETR performing better than UNETR, but both fell short of nnU-Net’s performance.

\begin{table}[!htbp]
\caption{Network architecture comparison for PCC segmentation on the test set}\label{tab:network_comparison}
\begin{tabular*}{\textwidth}{@{\extracolsep\fill}lccc}
\toprule
Network Architecture & DSC & NSD & F1 score \\
\midrule
UNETR & $0.718 \pm 0.146$ & $0.635 \pm 0.188$ & 0.646 \\
Swin UNETR     & $0.740 \pm 0.150$ & $0.680 \pm 0.172$ & 0.720 \\
nnU-Net     & $\mathbf{0.860 \pm 0.093}$ & $\mathbf{0.814 \pm 0.148}$ & \textbf{0.857} \\
\botrule
\end{tabular*}
\footnotetext{Results are based on experiments conducted with the TKA annotation.}
\end{table}

\subsection{Results by Genetic Subtype}

Table~\ref{tab:genetic_subtype_performance} presents the segmentation performance of the TKA and TB models across different genetic subtypes on the 30\% test data subset. For all genetic subtypes, the TKA model consistently outperformed TB on detection F1 score, DSC and NSD. The biggest performance gain was observed for the \textit{VHL/EPAS1} and sporadic genetic subtypes, where TKA achieved higher DSC (0.882 and 0.872), NSD (0.915 and 0.782), and F1 scores (0.941 and 1.000), respectively. 

\begin{table}[!htbp]
\caption{Performance by Genetic Subtypes on the test Set}\label{tab:genetic_subtype_performance}
\begin{tabular*}{\textwidth}{@{\extracolsep\fill}l l c c c}
\toprule
Gen Type & Method & DSC & NSD & F1 \\
\midrule
\multirow{2}{*}{\textit{SDHx}} 
  & TKA & \textbf{0.820 ± 0.148} & \textbf{0.762 ± 0.209} & \textbf{0.625} \\
  & TB  & 0.789 ± 0.129 & 0.724 ± 0.214 & 0.588 \\
\addlinespace
\multirow{2}{*}{\textit{VHL/EPAS1}} 
  & TKA & \textbf{0.882 ± 0.078} & \textbf{0.915 ± 0.072} & \textbf{0.941} \\
  & TB  & 0.828 ± 0.174 & 0.855 ± 0.197 & 0.667 \\
\addlinespace
\multirow{2}{*}{Kinase} 
  & TKA & \textbf{0.858 ± 0.092} & \textbf{0.832 ± 0.138} & \textbf{0.818} \\
  & TB  & 0.805 ± 0.145 & 0.746 ± 0.263 & 0.667 \\
\addlinespace
\multirow{2}{*}{Sporadic} 
  & TKA & \textbf{0.872 ± 0.066} & \textbf{0.782 ± 0.136} & \textbf{1.000} \\
  & TB  & 0.828 ± 0.144 & 0.724 ± 0.191 & 0.769 \\
\botrule
\end{tabular*}
\end{table}

\newpage
\section{Discussion and Conclusion}\label{sec5}

In this study, we introduced a robust deep learning-based framework utilizing nnU-Net for the accurate 3D segmentation of pheochromocytoma (PCC) tumors from contrast-enhanced abdominal CT scans. The key contribution of our approach lies in systematically exploring a diverse set of anatomical priors to enhance segmentation accuracy. Our results clearly demonstrate the critical role anatomical context plays in improving segmentation performance, emphasizing the contribution of certain organ-specific priors.

Among the eleven annotation strategies evaluated, the Tumor + Kidney + Aorta (TKA) approach yielded the highest segmentation performance across multiple metrics, including Dice Similarity Coefficient (DSC), Normalized Surface Distance (NSD), and instance-wise F1 score. In particular, compared to the previously used Tumor + Body (TB) strategy, TKA demonstrated statistically significant improvements—with a 25.84\% increase in F1 score at an IoU threshold of 0.5, higher DSC (\textit{p} = 0.0097), and superior NSD performance (\textit{p} = 0.0110), when evaluated on the test set. TKA’s consistent superiority over TB was further confirmed through five-fold cross-validation on the full dataset, highlighting its robustness and generalizability.

Our findings also highlight the importance of selecting appropriate anatomical priors for effective PCC segmentation. Annotations that included the kidney together with adjacent small organs (TKA, TKAG) or the kidney alone (TK) yielded stronger performance than other strategies, emphasizing the value of providing relevant spatial context. Conversely, strategies with an increased number of classes (e.g., TBLSKAG) and those incorporating large-volume organs—such as the liver and spleen (TLSA), which are substantially larger than typical PCC tumors—demonstrated reduced segmentation accuracy. This decline is likely driven by class imbalance during training and the added difficulty of learning to differentiate a broader set of anatomical structures. Importantly, the TB annotation, which relied on a broad body region prior without specific organ localization, showed reduced performance, potentially due to the lack of meaningful anatomical context for precise PCC tumor delineation. Moreover, training with tumor-only annotations was found to be insufficient for achieving optimal segmentation outcomes.

Additionally, our segmentation framework demonstrated accurate quantification of tumor burden, a vital clinical parameter in managing PCC. The high correlation (\( R^2 = 0.968 \)) between predicted and ground truth tumor volumes using the TKA annotation emphasizes its potential clinical utility, particularly in monitoring disease progression and evaluating treatment responses. A comparative analysis of network architectures further confirmed the superior performance of the nnU-Net framework over transformer-based models such as UNETR and Swin UNETR for this task.

In conclusion, the findings of our study highlight the critical role of carefully selecting anatomical priors in deep learning-based segmentation tasks. The proposed segmentation framework represents a substantial advancement in automated pheochromocytoma segmentation, offering the potential to improve clinical workflows, enable more reliable disease monitoring, and support enhanced prognostic assessments. Moreover, accurate PCC segmentation can serve as a valuable tool for predicting the underlying genetic cluster associated with the tumor, providing a cost-effective alternative to genetic testing for initial risk stratification. 

Despite the promising results of this study, certain limitations should be acknowledged. The dataset was retrospectively collected from a single institution, which may limit the generalizability of the findings to broader populations or imaging settings. Although five-fold cross-validation was performed to assess robustness, external validation using multi-center datasets with greater demographic and scanner variability would further support the clinical applicability of the proposed approach. Future work will focus on evaluating the generalizability of the proposed segmentation framework across larger, multi-institutional datasets and exploring its integration into broader clinical decision-support systems. Additionally, the framework could be extended to segment paragangliomas (non-adrenal neuroendocrine tumors), potentially incorporating additional imaging context, such as functional PET imaging, to better localize tumors that are often small, heterogeneous, and difficult to identify on CT alone.

\section*{Acknowledgements}
This work was supported by the Intramural Research Program of the NIH Clinical Center (project number 1Z01 CL040004). The research used the high-performance computing facilities of the NIH Biowulf cluster. The contributions of the NIH author(s) were made as part of their official duties as NIH federal employees, are in compliance with agency policy requirements, and are considered Works of the United States Government. However, the findings and conclusions presented in this paper are those of the author(s) and do not necessarily reflect the views of the NIH or the U.S. Department of Health and Human Services.


\newpage

\bibliography{sn-bibliography}

\end{document}